\documentclass[a4,twocolumn,prb,aps,showpacs,superscriptaddress,floatfix]{revtex4-1}
\usepackage[utf8]{inputenc}
\usepackage{bm}
\usepackage[usenames]{color}
\usepackage{multirow}
\usepackage{amssymb}
\usepackage{amsbsy}
\usepackage{amsmath}
\usepackage{stmaryrd}
\usepackage{graphicx}
\usepackage{epsfig}
\usepackage{placeins}
\usepackage{ulem}
\usepackage{hyperref}
\hypersetup{
colorlinks=true,
citecolor=blue,
linkcolor=red,
urlcolor=black
}

\usepackage{relsize}

\makeatletter

\renewcommand{\i}{\textrm{i}}

\newcommand{\ie}{{i.e.}}
\newcommand{\eg}{{e.g.}}

\newcommand{\be}{\begin{equation}}
\newcommand{\ba}{\begin{align}}
\newcommand{\ee}{\end{equation}}
\newcommand{\ea}{\end{align}}

\newcommand{\perpp}{\scriptscriptstyle{\perp}}
\newcommand{\parallell}{\scriptscriptstyle{\parallel}}

\newcommand{\p}{\scriptscriptstyle{+}}
\newcommand{\m}{\scriptscriptstyle{-}}
\newcommand{\Kp}{K_{\p}}
\newcommand{\Km}{K_{\m}}
\newcommand{\phip}{\phi_{\p}}
\newcommand{\phim}{\phi_{\m}}

\newcommand{\Jperp}{J_{\perpp}}
\newcommand{\LA}{L_{\textrm{\tiny A}}}
\newcommand{\LB}{L_{\textrm{\tiny B}}}
\newcommand{\SvN}{S_{\textrm{\tiny vN}}}
\newcommand{\jc}{j_{\textrm{\scriptsize c}}}
\newcommand{\rhoV}{\rho_{\textrm{\tiny V}}}
\newcommand{\lV}{l_{\textrm{\tiny V}}}
\newcommand{\gapM}{\Delta E_{\textrm{\tiny M}}}
\newcommand{\gapEx}{\Delta E_{\textrm{\scriptsize ex}}}
\newcommand{\rhoA}{\rho_{\textrm{\tiny A}}}

\newcommand{\crit}{\textrm{\scriptsize cr}}
\newcommand{\cdw}{\textrm{\tiny CDW}}

\newcommand{\sG}{\textrm{\tiny sG}}
\newcommand{\gs}{\textrm{\scriptsize gs}}
\newcommand{\ex}{\textrm{\scriptsize ex}}

\begin{document}
\title{Vortex and Meissner phases of strongly-interacting bosons on a two-leg ladder}

\author{M. Piraud} 
\author{F. Heidrich-Meisner}
\affiliation{Department of Physics and Arnold Sommerfeld Center for Theoretical Physics, Ludwig-Maximilians-Universit\"at M\"unchen, 80333 M\"unchen, Germany}
\author{I. P. McCulloch}
\affiliation{Centre for Engineered Quantum Systems, The University of Queensland, Brisbane, QLD 4072, Australia}
\author{S. Greschner}
\author{T. Vekua}
\affiliation{Institut f\"ur Theoretische Physik, Leibniz Universit\"at Hannover, 30167~Hannover, Germany}
\author{U. Schollw\"ock}
\affiliation{Department of Physics and Arnold Sommerfeld Center for Theoretical Physics, Ludwig-Maximilians-Universit\"at M\"unchen, 80333 M\"unchen, Germany}

\date{\today}

\begin{abstract}
We establish the phase diagram of the strongly-interacting Bose-Hubbard model defined on a two-leg ladder geometry in the presence  of a homogeneous flux. Our work is motivated by a recent experiment [Atala et al., Nature Phys. {\bf 10}, 588 (2014)], which studied the same system, in the complementary regime of weak interactions.
Based on extensive density matrix renormalization group simulations and a bosonization analysis, we 
fully explore the parameter space spanned by filling, inter-leg tunneling, and flux. As a main result, we demonstrate the existence of gapless and gapped Meissner and vortex phases, with the gapped states emerging in 
Mott-insulating regimes. We calculate experimentally accessible observables such as chiral currents and vortex patterns.
\end{abstract}

%\pacs{03.75.Lm, 05.30.Jp, 37.10.Jk}

%37.10.Jk 	Atoms in optical lattices
%03.75.Hh 	Static properties of condensates; thermodynamical, statistical, and structural properties
%05.30.Fk 	Fermion systems and electron gas
%05.30.Jp 	Boson systems
%67.85.Hj 	Bose-Einstein condensates in optical potentials
%03.75.Lm 	Tunneling, Josephson effect, Bose-Einstein condensates in periodic potentials, solitons, vortices, and topological excitations
%75.10.Jm 	Quantized spin models, including quantum spin frustration

\maketitle

{\it Introduction.}
The quantum states  of interacting electrons in the presence of spin-orbit coupling 
and magnetic fields are attracting significant attention in condensed matter physics
because of their connection to Quantum Hall physics~\cite{Thouless82}, topological insulators~\cite{kane05,hasan10,qi11}
and the emergence of unusual excitations in low dimensions~\cite{kitaev00,Fu08}.
Recent progress with quantum gas experiments has led to the realization of artificial gauge
fields~\cite{dalibard11}, both in the continuum~\cite{lin09,lin11,jimenez12} and for bosons in optical lattices~\cite{aidelsburger11,struck12,aidelsburger13,miyake13},
paving the way for 
future experiments 
on  the interplay of interactions, dimensionality, and gauge fields in a systematic manner.  This has motivated theoretical research into the physics of strongly interacting particles in the 
presence of abelian and non-abelian gauge fields
and various questions such as the Quantum Hall effect with bosons~\cite{sorensen05,palmer06,hafezi07,cooper08,fetter09,moeller09,senthil13,regnault13},
unusual quantum magnetism~\cite{cole12,radic12,cai12,orth13}, and the emergence of
topologically protected phases~\cite{grusdt13,grusdt14a,grusdt14b} have been addressed.

\begin{figure}
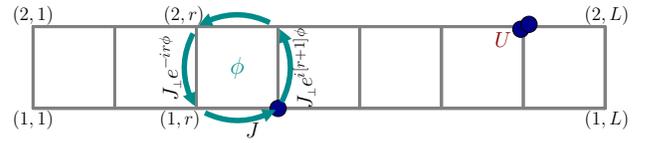

   \centering
  \includegraphics[width=0.95\columnwidth]{{{Model_pap2}}}
  \vspace{-0.01\textheight}
   \caption{(Color online) Sketch of the model Eq.~\eqref{eq:hamiltonian}: bosons on a two-leg ladder,
   with $J$ and $\Jperp$ the hopping matrix elements along the legs and
rungs, respectively, with $\phi$ the magnetic flux per plaquette, and $U$ the onsite interaction strength.}
\label{fig:sketch}
\end{figure}

Given the complicated interplay between interactions, gauge fields and dimensionality, one often has to resort to mean-field approaches to build up intuition for the emergent phases,
which should be complemented by reliable analytical and numerical results. 
In one dimension, both bosonization~\cite{giamarchi} and numerical  techniques such as the  density matrix renormalization
group (DMRG) method~\cite{white92,schollwoeck05,schollwoeck11} provide powerful tools to characterize the emergent quantum phases.
Here we consider  interacting bosons on a two-leg ladder in the presence of a homogeneous magnetic flux (see Fig.~\ref{fig:sketch} for a sketch of the model and definitions of parameters).
Such a system has been realized in a recent experiment with bosons in optical lattices~\cite{atala14},
yet in the weakly-interacting regime of high densities per site.
The existence of a transition between a phase with Meissner-like chiral currents and a vortex phase as a function of flux and rung tunneling strength has been demonstrated~\cite{atala14},
reminiscent of the field-dependence of currents in type-II superconductors.
Here we  provide complementary insights into the emergent phases in the 
strongly-interacting case where, in particular, also Mott-insulating phases can appear~\cite{vekua03,crepin11}. 

% ------------------------------- FIGURE 1 ----------------------------------
% ---------------------------------------------------------------------------

Bosons on a ladder subjected to gauge fields have been the topic of
previous theoretical work~\cite{orignac01,cha11,dhar12,dhar13,petrescu13,huegel14,wei14,tokuno14} (see also~\cite{lim08,moeller10} for 2D lattices), yet complete quantitative phase diagrams are lacking.
In our work, we use DMRG to systematically explore the full dependence on $\Jperp$, $\phi$, and filling and, as a main result, we observe both gapped and gapless Meissner and vortex phases 
for strongly-interacting bosons.
We focus on the gapped phases that emerge 
at a filling of one boson per rung, for which we present detailed results for chiral currents,
the vortex density and current patterns in the vortex phase.
In this Mott phase, Meissner currents are suppressed compared to superfluid phases, and can even decay to zero 
for an infinitely strong Hubbard interaction in the limit of large rung couplings $\Jperp \gg J$. 

% ------------------------------- FIGURE 2 ----------------------------------
\begin{figure}
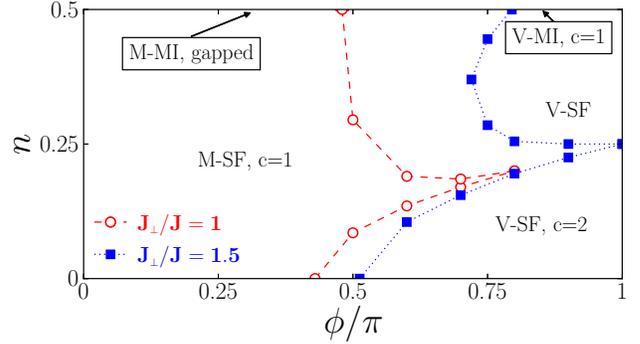

   \centering
   \includegraphics[width=0.95\columnwidth]{{{fig_phase-diag_pap2}}}
  \vspace{-0.01\textheight}
   \caption{(Color online) Phase diagram of HCBs for $\Jperp/J=1$ (circles) and $\Jperp/J=1.5$ (squares) 
   as a function of flux $\phi$ and density $n$ (DMRG data, $L=101$).
 The region $0.5<n\leq 1$ is related to the low-density regime by particle-hole symmetry.}\label{fig:diag-phase}
\end{figure}
% ---------------------------------------------------------------------------
{\it Hamiltonian and observables.}
The Hamiltonian is given by (see Fig.~\ref{fig:sketch}): 
\begin{eqnarray}
H &=& \sum_{\ell =1,2;r=1}^L \left\lbrack -J \left( a^\dagger_{\ell,r+1} a_{\ell,r} + \textrm{H.c.} \right)
   + \frac{U}{2} n_{\ell,r} (n_{\ell,r} -1)\right\rbrack \nonumber \\
   && -\Jperp  \sum_{r=1}^L \left( e^{-i r \phi} a^\dagger_{1,r} a_{2,r} + \textrm{H.c.} \right)
\label{eq:hamiltonian}
\end{eqnarray}
on a ladder with $L$ rungs where $a^\dagger_{\ell,r}$ creates a boson on site $\ell=1,2$ of the $r$th rung.
Energy is measured in units of $J$. 
We define the filling as $n=  N/(2L)$, where $N$ is the total number of bosons.

On the one hand, the Hamiltonian~\eqref{eq:hamiltonian} can be viewed as a minimal model for describing the edge states of a two-dimensional interacting Bose system 
 pierced by a flux.
On the other hand, we can interpret the system as a  one-dimensional two-component gas~\cite{petrescu13,huegel14}, where the two species are labeled with $\ell=1,2$.
In the latter case, the term proportional to $\Jperp$ breaks the $U(1)$ symmetry related to the conservation of the particle numbers of the
individual components.

Local currents will be a key quantity for characterizing different phases. We define the currents along the legs  $j^{\parallell}_{\ell,r}$ and rungs $j^{\perpp}_{r}$ as
\begin{align}
j^{\parallell}_{\ell,r} &= i J \left( a^\dagger_{\ell,r+1} a_{\ell,r} - a^\dagger_{\ell,r} a_{\ell,r+1} \right)
\label{eq:CurrentsDefs}\\
j^{\perpp}_{r} &= i \Jperp \left( e^{-i r \phi} a^\dagger_{1,r} a_{2,r} - e^{i r \phi} a^\dagger_{2,r} a_{1,r} \right)\,.
\label{eq:CurrentsDefs2}
\end{align}
The chiral (or Meissner) current is
$ \jc = \partial {\cal E}_0/ \partial \phi=\frac{1}{2L} \sum_{r} \langle j^{\parallell}_{1,r}-j^{\parallell}_{2,r} \rangle$, 
where ${\cal E}_0$ is the ground-state energy per site.
Note that the operators given in Eqs.~\eqref{eq:CurrentsDefs}-\eqref{eq:CurrentsDefs2} depend on the gauge, but the associated expectation values are gauge invariant~\cite{moeller10}, as can be explicitly seen in the definition of the Meissner current.
For the data shown in the figures,  $\jc$ is computed by restricting the sum to  $r \in [-L/4, L/4]$ to suppress 
boundary effects, since in DMRG simulations we use open boundary conditions.

% ------------------------------- FIGURE 3 ----------------------------------
\begin{figure}
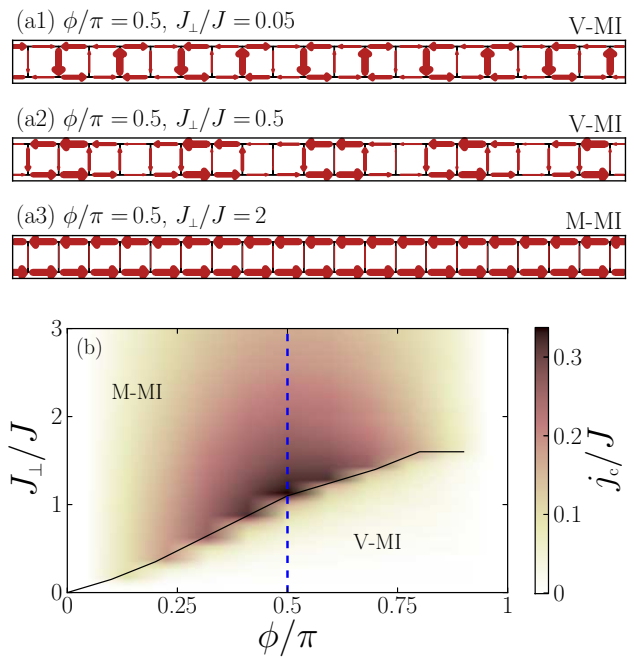

   \includegraphics[width=0.95\columnwidth]{{{Beau-currents_pap}}}\\
  \vspace{0.02\textheight}
\includegraphics[width=0.95\columnwidth]{{{fig_currents_map_pap}}}
  \vspace{-0.01\textheight}
   \caption{(Color online)
    (a1)-(a3) Typical current patterns  for $n=0.5$, $\phi/\pi=0.5$ and $\Jperp/J=0.05$, 0.5 and 2
    and (b) chiral current $j_{c}$
    as a function of $\phi$ and $\Jperp$ (HCBs, $L=101$). The width of the arrows in (a1)-(a3) is proportional to the expectation values of the local currents. In (b), the solid line locates the maximum of $\jc=\jc(\phi)$ at fixed $\phi$
    and the dashed line the cut considered in Fig.~\ref{fig:n05_currents_cuts}.
    }
     \label{fig:n05_currents}
\end{figure}
% ---------------------------------------------------------------------------

{\it Phase diagram as a function of filling.}  
Let us start by giving an account of our main results, which can be inferred from considering the limit of {hard-core bosons} (HCBs), i.e., $U/J=\infty$.
Figure~\ref{fig:diag-phase} shows the phase diagram for this case as a function of  $n$ and $\phi$ for $\Jperp/J=1$ and 1.5.
These results are  based on a combination of a field-theory analysis
and DMRG simulations for current correlation functions, the von Neumann entropy, excitation gaps, and the 
equation of state $n=n(\mu)$, where $\mu$ is the chemical potential.

In Fig.~\ref{fig:diag-phase} we identify mainly four types of phases.
At half-filling ($n=0.5$), there is a Mott insulator (MI) with a mass gap for any value of $\phi$ and $\Jperp\not= 0$.
At small values of $\phi$, we find a Meissner phase (M-MI)
while at large $\phi$, a gapless vortex state exists (V-MI).
This confirms the prediction of a Mott gap for HCBs at $n=0.5$ and $\Jperp \neq 0$~\cite{vekua03,crepin11} and the emergence of the Meissner currents and a vortex phase for $\phi \neq 0$~\cite{petrescu13}.
At finite values of $U/J<\infty$, there will be a MI-SF transition, with critical interaction strength depending on $\Jperp/J$ \cite{suppmat}.
At  $n<0.5$, there are superfluid phases which can again be divided into a Meissner superfluid (M-SF) and a vortex superfluid (V-SF). 
The terms Meissner and vortex state are justified by the existence of characteristic current patterns.
Examples for $n=0.5$ are shown in Figs.~\ref{fig:n05_currents}(a1)-(a2) (V-MI) and Fig.~\ref{fig:n05_currents}(a3) (M-MI) (current patterns in the M-SF and V-SF are qualitatively similar to the ones in the M-MI and V-MI, respectively: see Figs.~S4(a)-(c)~\cite{suppmat}). 
The Meissner phases have vanishing rung currents $\langle j_r^{\perpp}\rangle$ but a finite chiral current $\jc$, 
while in the vortex phase,  $\langle j_r^{\perpp}\rangle\not =0$ on finite systems, with various possible vortex patterns.
The M-SF phase has one gapless mode (central charge $c=1$), while the V-SF has $c=2$.
We expect M-SF and  V-SF to be  adiabatically connected to the corresponding
phases established at weak interactions~\cite{atala14,tokuno14,orignac01}.

The M-SF phase penetrates into the V-SF phase at intermediate values of $\Jperp \sim J$. 
The vicinity of $\phi=\pi$ is special because at $n=0.25$, a gapped  charge-density-wave (CDW) phase emerges at $\Jperp \gtrsim 1.3J$. Once this happens, the M-SF phase touches this phase, splitting the V-SF into two lobes.
Eventually,
both the V-MI and the upper lobe of the V-SF phase disappear for large $\Jperp \gtrsim 1.7J$. 
For $\Jperp \gtrsim 1.5 J$, we find a jump in density at $\phi=\pi$, from  $n>0.25$  to the gapped $n=0.5$ state,
which for $\Jperp/J\to \infty$ extends down to $n=0.25$.

{\it Effective field theory.}
The nature of the phase transitions can be elucidated using bosonization.
If we fix $\Jperp\neq 0$ and change the flux at half-filling,
 there is a commensurate-incommensurate (C-IC) quantum phase transition~\cite{giamarchi} from  
a gapped ($\phi<\phi^{\crit}$) to a gapless ($\phi>\phi^{\crit}$) behavior of the relative phase fluctuations  of the two-leg system, 
whereas the total density mode is always gapped for strong interactions~\cite{suppmat}.
Away from $n=0.5$, the total density mode becomes immediately gapless~\cite{crepin11} and
there is a C-IC transition in the relative degrees of freedom from a gapped to a gapless behavior as a function of flux~\cite{orignac01}.
This picture is confirmed by DMRG results for the von Neumann entropy (see Figs.~S3 and S7~\cite{suppmat}) and consistent with the 
 transitions shown in Fig.~\ref{fig:diag-phase}.

The emergence of a two-component Luttinger liquid (LL) at large values of $\phi$ becomes transparent in the low-density limit where it is connected with the development of a double-minimum structure in the single-particle dispersion $\epsilon_k$ for $\phi>\phi^{\crit}(\Jperp)$~\cite{huegel14,tokuno14}.
Note that the physics at low densities is very similar to that of frustrated chains in high magnetic fields below saturation (see~\cite{arlego12,kolezhuk12,shyiko13} and references therein).
For bosons and in the limit of vanishing density, once the single-particle dispersion acquires a double-minimum, the $c=2$ LL is stabilized.
To show this, we solve the low-energy scattering problem of two bosons and extract the relevant scattering lengths.
There are two important scattering processes at low energy: either the two bosons belong to the same minimum of $\epsilon_k$ (intra-species scattering) or they belong to different minima (inter-species scattering).
In 1D, the  scattering length is related to the  scattering phase shift via
$a_{i,j}=\lim_{K\to 0} \big[\cot(\delta_{i,j})/K\big]$, where $K$ is the  relative momentum of the two bosons and $i,j=1,2$ distinguish bosons belonging to the minimum in $\epsilon_k$ at $k<0$ or $k>0$, respectively.
The scattering length is related to the amplitude of the contact potential of the two-component Bose gas $U_{i,j}(x-x')=g_{i,j}\delta(x-x')$ with $g_{i,j}=-2/(a_{i,j}m) $.
By comparing the scattering lengths $a_{i,j}$ to each other we find that in strong coupling $a_{1,1}>a_{1,2}$, such that once the double-minimum structure appears in $\epsilon_k$, the $c=2$ LL is energetically preferred for $n\to 0$, consistently with the mean-field argument of~\cite{tokuno14} and with the DMRG results shown in Fig.~\ref{fig:diag-phase}.

{\it Large $\Jperp/J$ limit.}
Another interesting limit amenable to an analytical treatment is the case of strong rung tunneling $\Jperp/J\to \infty$.
In that regime we introduce a pseudo-spin-$1/2$ operator $\vec{S}_{r}$ on each rung $r$ associated to the states 
$(|1,0 \rangle_r + e^{i r\phi }|0,1 \rangle_r)/\sqrt{2}\to|\downarrow \rangle_r$, and $|0,0 \rangle_r  \to|\uparrow \rangle_r $. The effective spin-$1/2$ model for the special
case of $\phi=\pi$ and to first order in $J^2/\Jperp$ is~\cite{suppmat}:
\begin{equation}
\label{effectivespinhalf}
H_{\frac{1}{2}} = \frac{J^2}{2|\Jperp|} \sum_{r} \left( 2 S^z_{r} S^z_{r+1} - \Big[ S^{\p}_{r} (\frac{1}{2} - S^z_{r+1}) S^{\m}_{r+2} + 
{\rm h.c.} \Big] \right)\,.
\end{equation}
In this basis, $n=0.5$ corresponds to the fully 
polarized state $\prod_r | \downarrow \rangle_r $ and the vacuum of bosons $n=0$  corresponds to $\prod_r | \uparrow \rangle_r $,
while $n=0.25$ implies a vanishing magnetization $\langle S^z_{r}\rangle $.
 The classical N\'{e}el-state
$\left|...\uparrow\downarrow\uparrow\downarrow...\right\rangle$ is an
eigenstate of the effective model Eq.~\eqref{effectivespinhalf} and for quarter-filling it becomes the
ground state due to the dominant Ising interaction.
Hence, in the vicinity of $\phi=\pi$ the ground state of bosons for $\Jperp/J\gg 1$ at quarter-filling ($n=0.25$) is a doubly-degenerate CDW state, which breaks translational invariance. Away from  $\phi\sim\pi$, the  effective model undergoes a Kosterlitz-Thouless transition at some $\phi^{\crit}_{\cdw}$ from the N\'eel state ($\phi^{\crit}_{\cdw}<\phi\le \pi$) into a gapless XY phase ($\phi\le \phi^{\crit}_{\cdw}$), the latter being characterized by $c=1$.
The existence of a fully gapped CDW state at $n=0.25$ for strong $\Jperp/J$ in the vicinity of $\phi=\pi$ and of
a  direct transition from the fully gapped state to a $c=1$ phase with decreasing $\phi$ explains the tendency of the M-SF  to  pierce the V-SF (see Fig.~\ref{fig:diag-phase}). 

The effective spin-$\frac{1}{2}$ model Eq.~\eqref{effectivespinhalf} further  unveils the presence of a 
metamagnetic behavior just below the saturation magnetization, corresponding  to a jump in the density of bosons from $n=0.25$ to $n=0.5$ at  $\Jperp/J\to \infty$. 
Due to the absence of spin-inversion symmetry in Eq.~(\ref{effectivespinhalf}) there is no such jump from $n=0.25$ to $n=0$.
For $\Jperp/J<\infty$, this metamagnetic behavior survives with a jump between some $n> 0.25$ to $n=0.5$, which explains the numerical data shown in Figs.~S1 and S2~\cite{suppmat}.
 
% ------------------------------- FIGURE 4 ----------------------------------
\begin{figure}
\includegraphics[width=0.95\columnwidth]{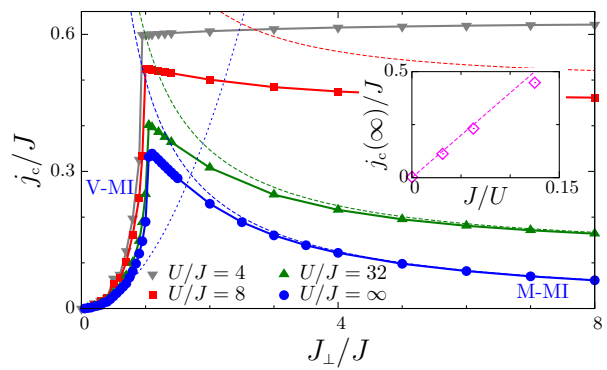}
  \vspace{-0.01\textheight}
   \caption{(Color online)
   Cut through the phase diagram Fig.~\ref{fig:n05_currents}(b) at $\phi/\pi=0.5$ for HCBs
   as well as $U/J=4$, 8 and 32.
   Dashed lines:  Theoretical predictions for $\Jperp\ll J$ and $\Jperp \gg J$ [see Eqs.~\eqref{eq:weakK} and \eqref{eq:largeK}].
   Inset: Asymptotic value $\jc(\Jperp/J \rightarrow \infty)$ as a function of $1/U$, 
   together with $\jc(\infty)=4J^2/U$.
   ($U/J < \infty $: $L=60$, $L= 201$ for $U/J=\infty$).
   } 
   \label{fig:n05_currents_cuts}
\end{figure}
% ---------------------------------------------------------------------------
% ------------------------------- FIGURE 5 ----------------------------------
\begin{figure}
\includegraphics[width=0.95\columnwidth]{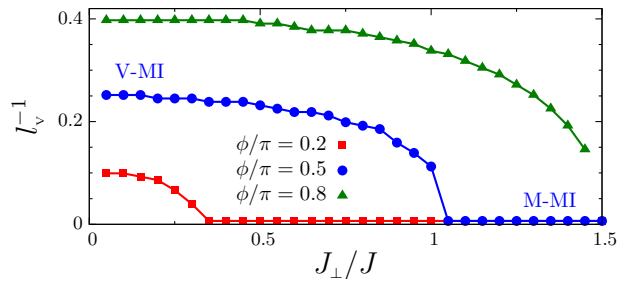}
  \vspace{-0.01\textheight}
   \caption{(Color online) Vortex density $\lV^{-1}$, i.e.,  inverse typical extension $\lV$ of the vortices (in lattice sites),
   versus $\Jperp$, for $\phi/\pi=0.2$, 0.5 and 0.8 ($L=101$). 
    }
   \label{fig:vphase}
\end{figure}
% ---------------------------------------------------------------------------

{\it Dependence of currents on $\phi$ and $\Jperp$.}
Figure~\ref{fig:n05_currents}(b) shows the chiral current as a function of $\phi$ and $\Jperp/J$ for HCBs at $n=0.5$.
The chiral current takes a maximum at the transition from the V-MI phase to the M-MI phase.
Using field theory, we derive an expression for the chiral current, in the regime $\Jperp\ll J$ and for small $\phi$
\begin{equation}
\jc  \sim \frac{ \Jperp^2}{ J \phi^{3-1/K_0}}+O(\Jperp^4)\, ,
\label{eq:weakK}
\end{equation}
where  $K_0$ is the LL parameter for the Bose-Hubbard model of
decoupled chains ($\Jperp=0$), and ranges from $K_0=\infty$, for $U=0$, to $K_0=1$, for HCBs.
The $\jc \propto \Jperp^2$ behavior is a generic result, valid for any repulsion $U$ and filling~\cite{suppmat}.
Equation~(\ref{eq:weakK}) implies that $\jc$ increases the fastest with $\Jperp$ at {\it small} values of $\phi$.
In particular, for HCBs, we obtain $\jc \sim (\Jperp/\phi)^2$.

For the opposite limit of large $\Jperp\gg J$, we use perturbation theory at $n=0.5$~\cite{suppmat} to derive that  for $U/J \gg 1$
\begin{equation}
\jc = \frac{J^2 (4 \Jperp+U)^2}{2 \Jperp U (2
\Jperp+U)}  \sin(\phi)\,. \label{eq:largeK}
\end{equation}
Therefore, in the limit of infinitely strong interactions, the chiral current decays to zero in the M-MI as $\jc \propto 1/\Jperp$, contrary to the behavior at finite $U/J<\infty$ where the chiral current
saturates at large $\Jperp \gg J$, as $\jc(\infty) \propto 1/U$ (see the inset in Fig.~\ref{fig:n05_currents_cuts}).
This latter saturation is known from the $U=0$ limit~\cite{huegel14,atala14} and is also observed in M-SF phases for $U \neq 0$ (results not shown).

Figure~\ref{fig:n05_currents_cuts} presents a cut of Fig.~\ref{fig:n05_currents} at $\phi=\pi/2$, together with finite $U/J$ data.
The analytical predictions for the weak- and strong-coupling regimes from Eqs.~\eqref{eq:weakK} and \eqref{eq:largeK}
agree very well with our DMRG data for $U/J \gg1$ [dashed lines in Fig.~\ref{fig:n05_currents_cuts}].
The essential features of the HCB case carry over to finite values of $U/J<\infty$.
A finite $U$ suppresses the chiral current compared to $U=0$, which should be accessible in experiments.

The vortex phases can be further characterized by their current patterns which bear well-defined structures, with varying spatial extension and density as a function of $\Jperp$ and $\phi$.
For the parameters of Fig.~\ref{fig:n05_currents}(a1), the sign of the current alternates along the legs, reminiscent of the chiral MI phase discussed in~\cite{dhar12,dhar13} 
These structures can be quantitatively studied by analyzing the rung currents $\langle j_{r}^{\perpp}\rangle$.
Figure~\ref{fig:vphase} shows the vortex density $\lV^{-1}$ at $n=0.5$ as a function of $\Jperp/J$ for various values of $\phi$, where $\lV$ is the typical size of 
vortices extracted from the Fourier transform of the real-space patterns $\langle j_{r}^{\perpp}\rangle$ over $r \in [-L/4, L/4]$.
This can be interpreted as a measure of the order parameter of the transition from the Meissner into the vortex phase \cite{orignac01}.
As expected, $\lV^{-1}$ decreases to zero as the transition into the M-MI phase is approached, where only longitudinal currents survive. This is consistent with field theory predictions, which also provide that in the $\Jperp \ll J \phi$ limit, $\lV^{-1} \sim \phi$~\cite{suppmat}. 
The rung-current correlation function $\langle j_r^{\perpp} j_{r'}^{\perpp} \rangle $ decays algebraically  in all vortex phases (see Fig.~S5~\cite{suppmat}), unlike in
the so-called chiral MI phase~\cite{dhar12,dhar13} realized for $U/J<\infty$, $\phi=\pi$, $\Jperp=J$, and $n=1$, which  has long-range rung-current correlations.

{\it Summary.}
Based on a combined DMRG and field-theoretical study, we obtained the phase diagram of strongly interacting bosons on a two-leg ladder in the presence of a homogeneous flux per plaquette.
We demonstrated the existence of both gapless and gapped Meissner and vortex phases, where the gapped Meissner phase emerges in the Mott-insulating regime.
The chiral current is suppressed by interactions and for HCBs it decays to zero  in the M-MI, with increasing $\Jperp$.
These results substantially extend previous studies of related models~\cite{dhar12,dhar13,petrescu13}  and confirm various predictions from field theory~\cite{orignac01,tokuno14}.
We provided analytical results for the weak- and strong-coupling limit, in very good agreement with numerical data. 
Our findings will provide guidance for future experimental studies (similar to \cite{atala14})
of the strongly-interacting regime.
The interaction strength, density and the ratio of hopping matrix elements can routinely be tuned in optical lattice experiment \cite{bloch08}, and so far, $\phi=\pi/2$~\cite{aidelsburger13,atala14} and $\phi=\pi$~\cite{miyake13} has been realized.
 Interesting extensions of our present study include the current patterns in harmonic traps. For this case, our results for $n=n(\mu)$ provide information about the real-space density profiles via the local density approximation.
Moreover, there is the possibility to stabilize vortex solids \cite{orignac01}, 
 which  are so far elusive in the 
strongly-interacting regime at incommensurate fillings. 
In the strong-coupling limit $U\gg J$,  vortex solids are not observed in our numerical
data  either in the superfluid or in the $n=0.5$ Mott phase, as opposed to the $n=1$ Mott phase 
for moderate values of $U/J$ \cite{dhar12,dhar13}, where a vortex 
solid appears at $\phi=\pi$.

{\it Note added.} Very recently, two more experimental studies have investigated fermions \cite{mancini15} and bosons \cite{leblanc15}
on ladders in optical lattices in the presence of artificial gauge fields.

We thank A. Paramekanti and I. Bloch for helpful discussions.
The research of M.P. was supported by the European Union through the Marie-Curie grant 'ToPOL' (No. 624033) (funded within FP7-MC-IEF).
This work was also supported in part by National Science Foundation Grant No. PHYS-1066293 and the hospitality of the Aspen Center for Physics.
S.G. and T.V. are supported by the QUEST (Center for Quantum Engineering and Space-Time Research) and DFG Research Training Group (Graduiertenkolleg) 1729,
and I.MC. acknowledges funding from the Australian Research Council Centre of Excellence for Engineered Quantum Systems.

\newpage
%\appendix
\renewcommand{\theequation}{S\arabic{equation}}
\setcounter{equation}{0}
\renewcommand{\thefigure}{S\arabic{figure}}
\setcounter{figure}{0}
\renewcommand{\thetable}{S\arabic{table}}
\renewcommand{\thesection}{S\arabic{section}}
\setcounter{section}{0}

\onecolumngrid  

\begin{center}
\textbf{\large{Supplemental Material for 'Vortex and Meissner phases of strongly-interacting bosons on a two-leg ladder'}}
\end{center}

\twocolumngrid

\section{Effective Field Theory for $\Jperp \ll J$}
\label{sec:efftheory}

For finite densities and weak rung tunneling $\Jperp$ we apply an effective field theory  \cite{giamarchi}, 
with the help of which we map out the ground-state phase diagram.
We introduce two pairs of conjugate bosonic fields ($\theta_{\ell},\phi_{\ell}$), for $\ell=1,2$,
describing, respectively, phase and density fluctuations of bosons on leg $\ell$, with commutation relations $[\theta_{\ell}(x),\partial_y\phi_{\ell'}]=i\delta_{\ell,\ell'}\delta(x-y)$.
The low-energy properties of the model given in Eq.~(1) of the main text are then governed by the following Hamiltonian density
\begin{align}
\label{EffHam}
{\mathcal H}(x)=&\frac{ v_{\p}}{2}\left[ \frac{(\partial_x \phip)^2}{ \Kp}+ \Kp(\partial_x \theta_{+})^2  \right]  
	     \\ &+\frac{ v_{\m}}{2}\left[ \frac{(\partial_x \phim)^2}{ \Km}+ \Km(\partial_x \theta_{\m}-\frac{\phi}{\sqrt{2\pi}})^2  \right] \nonumber \\
&-\cos {\sqrt{2\pi} \theta_{\m}} \sum_{m=0,1}  \lambda_{m}\cos{[m\sqrt{8\pi} \phip +4m\pi nx]}\nonumber
\end{align}
where  $\phi_\pm=(\phi_1\pm \phi_2)/\sqrt{2}$, $\theta_+=(\theta_1+ \theta_2)/\sqrt{2}$, $\theta_-=(\theta_1- \theta_2 +\phi x/\sqrt{\pi})/\sqrt{2}$,
and couplings constants $\lambda_m\sim \Jperp$;
$K_{\pm}$ are Luttinger-liquid parameters corresponding to the total and relative fluctuations of the two-leg ladder and $v_{\pm}$ are the corresponding velocities.
For $\Jperp \ll J$, $K_{\pm}=K_0(1-O(\Jperp^2))$ and $v_{\pm} =v_0+O(\Jperp^2)$ where $K_0$ and $v_0$ are the Luttinger-liquid parameter and sound velocity for the one-dimensional Bose-Hubbard model, respectively.
In particular, $K_0=\infty$ for $U/[Jn]\to 0$ and $K_0=1$ for $U/[Jn]\to \infty$.
One also has $v_0=\alpha J a$, where we fix the lattice constant $a=1$ in the following, 
where the proportionality constant $\alpha$ ranges from $\alpha \sim U/J$, for $U/[Jn]\to 0$, to $\alpha=\sin (\pi n)$ for $U/[Jn] \to \infty$.

The most important term in Eq.~(\ref{EffHam}) is the one proportional to $\lambda_0$, 
which, for small values of the flux and at any filling and interaction strength $U$, opens a gap, for arbitrarily small rung tunneling $\Jperp$,
in the antisymmetric sector. This gap is given by  $\Delta_{\m}\sim \Jperp^{1/(2-1/(2\Km))}$ for $\Jperp\ll J$ and 
the interaction term pins $\langle \theta_{\m}\rangle$, i.e., it  locks the relative phase of bosons on the two legs 
as long as $\phi<\phi^{\crit}$, where $\phi^{\crit}$ is determined by a (soliton) gap of the antisymmetric sector.
For the case of half-filling ($n=0.5$), and to unveil the role of the commensurate inter-sector interaction 
term $\lambda_1$, we apply a mean-field like decoupling that is justified due to the strongly relevant $\lambda_0$ 
coupling~\cite{vekua04,crepin11}.
We obtain an exactly solvable effective field theory, which is a direct sum of two quantum sine-Gordon models 
\be
{\mathcal H}= {\mathcal H}^{\m}_{\sG}+{\mathcal H}^{\p}_{\sG},
\ee
where
 \begin{align}
 {\mathcal H}^{\m}_{\sG}(x)=& \frac{ v_{\m}}{2}\left[(\partial_x \phim)^2+ (\partial_x \theta_{\m})^2  \right] -2\lambda \cos{\beta \theta_{\m}}\nonumber \\
& -\frac{v_{\m} \beta }{2\pi}A \partial_x \theta_{\m}+\frac{v_{\m}\Km}{4\pi} \phi^2, \nonumber
 \end{align}
 with $2\lambda=\lambda_0+\lambda_1 \langle \cos{\tilde \beta \phip} \rangle_{{\mathcal H}^{\p}_{\sG}} $,
 and 
 $${\mathcal H}^{\p}_{\sG}(x)=   \frac{ v_{\p}}{2}\left[ (\partial_x \phip)^2 + (\partial_x \theta_{\p})^2  \right] -2\tilde \lambda\cos{\tilde \beta \phip},$$ 
 with $2\tilde \lambda=\lambda_1 \langle \cos{\beta \theta_{\m}} \rangle_{{\mathcal H}^{\m}_{\sG}}$.
  In the previous equation, the short-hand notations $A= \sqrt{2\pi} \Km \phi$, $\beta=\sqrt{2\pi/\Km}$ and $\tilde \beta=\sqrt{8 \pi \Kp}$  were introduced.

The expectation values $\langle \cos{\tilde \beta \phip} \rangle_{{\mathcal H}^{\p}_{\sG}}$ and $\langle \cos{\beta \theta_{\m}} \rangle_{{\mathcal H}^{\m}_{\sG}}$ can be  evaluated in the vacuum of the quantum sine-Gordon models in an exact way, 
including  the phase with $\phi>\phi^{\crit}$, where the antisymmetric sector contains a finite density of topological solitons (vortices) in the ground state and becomes gapless (equivalently as for the symmetric sector in the case of a
finite doping away from half-filling). The vacuum energy density  ${\cal E}_0(\lambda,\phi)$ of the quantum sine-Gordon model is known exactly  \cite{Zamolodchikov93,Zamolodchikov95}, and the desired expectation values can be obtained from the 
Hellmann-Feynman theorem as $\partial_\lambda {\cal E}_0(\lambda,\phi)$.
 At half-filling and in the hard-core limit, $\Kp= 1-O(\Jperp^2)$ and a 
Kosterlitz-Thouless (KT) type renormalization group analysis of the marginal perturbation (i.e., the term proportional to 
$\tilde \lambda$) shows that any rung tunneling opens a Mott gap in the symmetric sector~\cite{vekua04,crepin11}. The 
gap is exponentially small in $\Jperp$, for $\Jperp\ll J$.  For finite values of $U/J<\infty$,
there exists a critical value of $\Jperp$ and a Mott gap opens 
for $\Jperp>J_{\perpp}^{\crit}(U)$ via a KT phase transition at $\Jperp=J_{\perpp}^{\crit}(U)>0$ for $U/J<\infty$.

First, we consider a fixed value of the flux and discuss the limit $\Jperp \ll J\phi$, where the 
exact ground-state energy can be expanded into a perturbation series, pressumably even with a 
finite convergence radius \cite{Zamolodchikov95} as
 ${\cal E}_0=-\frac{v_{\m}A^2}{\pi} \kappa (A,\lambda)$, where $\kappa(A ,\lambda)=\kappa(\xi)=\sum_{n=1}^{\infty}\kappa_n\xi^{2n}$, with $\xi=\lambda/(v_{\m}A^{2/(p+1)})$ and $p=\beta^2/(8\pi-\beta^2)$. 
The expressions for the coefficients $\kappa_n$ are known \cite{Zamolodchikov95}.
To leading order in $\Jperp/J$, the vacuum energy is
\be
\label{Zamseries}
{\cal E}_0(A,\lambda)=-\frac{v_{\m}A^2}{\pi}\left\{ \kappa_1 \frac{\lambda^2/v_{\m}^2}{
A^{(8\pi-\beta^2)/2\pi}} +O(\xi^4)\right\}
\ee
where
\be
\kappa_1=\pi^2 \left(\frac{2p}{p+1}\right)^{2(p-1)/(p+1)}\frac{\Gamma\left(\frac{1-p}{1+p}\right)}{\Gamma\left(\frac{2p}{p+1}\right)}
\ee
was calculated by Zamolodchikov \cite{Zamolodchikov95}.
In the hard-core limit, $\kappa_1=2\pi^2$ and it
increases with decreasing $U/J$.
The  dependence of the chiral current on $\Jperp$ is given by
\be
\label{chircur}
\jc=  \frac{\partial {\cal E}_0(\phi,\lambda) }{\partial \phi}\simeq \frac{\kappa_1 \Jperp^2}{J  \phi^{3-1/\Km}}+O(\Jperp^4).
\ee
Note that the proportionality to 
$\Jperp^2$, for $\Jperp\to 0$ and $\phi>0$, is completely generic, valid for any repulsion $U$ and filling. The vortex density, defined as the density of the phase slips (solitons) in the sine-Gordon model describing the relative phase fluctuations is given by $\rhoV \to   \langle\partial_x \theta_{\m}/\sqrt{\pi}\rangle_{H^{\m}_{\sG}}$ and in the limit $\Jperp \ll J\phi$, we have $\rhoV\sim \phi$.  

Next we consider the case  of a fixed rung tunneling $\Jperp\neq 0$ and elucidate the dependence on flux. At half filling 
we obtain the following picture: with increasing flux, 
at $\phi=\phi^{\crit}$ the antisymmetric sector undergoes a commensurate-incommensurate (C-IC) quantum phase transition \cite{giamarchi} from a gapped ($\phi<\phi^{\crit}$) to a gapless ($\phi>\phi^{\crit}$) behavior, whereas the 
symmetric sector always remains gapped since $\tilde \lambda=\lambda_1\partial {\cal E}_0(\phi,\lambda)/\partial \lambda \neq 0$, even if $\phi \gg \phi^{\crit}$, which can be seen from Eq.~(\ref{Zamseries}). Therefore, at half-filling the  Mott state is stable and increasing flux induces a C-IC quantum phase transition in the antisymmetric sector, from a fully 
gapped Meissner-Mott to a partially gapped Vortex-Mott phase.
Note that $\partial {\cal E}_0(\phi,\lambda)/\partial \lambda= \partial {\cal E}_0(0,\lambda)/\partial \lambda$ is independent of $\phi$ for $\phi<\phi^{\crit}$.
Hence the Mott gap is independent of $\phi$ for $\phi<\phi^{\crit}$.
 For $\phi>\phi^{\crit}$, $\partial {\cal E}_0(\phi,\lambda)/\partial \lambda$
continuously decreases with increasing $\phi$ in the V-MI~\cite{Zamolodchikov95}, starting from its Meissner-Mott value $\partial {\cal E}_0(0,\lambda)/\partial \lambda$. The Mott state at $\phi<\phi^{\crit}$ is similar to the rung-triplet phase \cite{vekua04} and can,
especially for strong $\Jperp$, be mimicked as a direct product of $\prod_r(|1,0 \rangle_r + e^{i r \phi} |0,1 \rangle_r)/\sqrt{2}$.

Away from half filling, the symmetric sector immediately becomes incommensurate and hence gapless.
In addition, the value of $\phi^{\crit}$ increases due to the weakened response in the coupling constant $\lambda$ from the symmetric sector. 
Hence, away from half filling, there is a C-IC transition in the 
antisymmetric sector from a gapped to a gapless behavior with increasing flux, with the 
symmetric sector providing an overall gapless background \cite{orignac01}. This describes the transition from a Meissner superfluid (M-SF) to a vortex superfluid (V-SF) state at incommensurate fillings.

The vortex density is $\rhoV=0$ for $\phi<\phi^{\crit}$ and beyond the  C-IC phase transition at $\phi =\phi^{\crit}$ upon further increasing the flux,
 $\rhoV$ increases with a square-root behavior which is characteristic for the C-IC transition, 
namely $\rhoV \sim \Theta(\phi-\phi^{\crit}) \sqrt{ \phi-\phi^{\crit} }$.
The chiral current, given in Eq.~(\ref{chircur}) behaves as described in \cite{orignac01}, namely, it increases linearly with flux until $\phi< \phi^{\crit}$ and then decreases in the vortex phase, consistent with the DMRG data shown in Fig.~\ref{fig:curr_cuts_suppmat}(b) for small rung tunneling.
Note that  $\jc=v_{\m}\Km\phi/(2\pi)$ for $\phi<\phi^{\crit}$ (where the vacuum of the
sine-Gordon theory does not contain solitons). This is the behavior in the  Meissner phases (both the M-SF and M-MI) while in the 
vortex phases (V-SF and V-MI) for $\phi>\phi^{\crit}$, $\jc-v_{\m}\Km\phi/(2\pi)\sim- \sqrt{ \phi-\phi^{\crit} }$.

The rung current is $\langle j^{\perpp} \rangle\to \langle \sin {2\pi \theta_{\m}} \rangle_{H^{\m}_{\sG}}$ and it 
is pinned at zero in the soliton-free vaccum of the sine-Gordon model 
($\phi<\phi^{\crit}$). The rung-current correlation function $ \langle j^{\perpp}(x)j^{\perpp}(y)  \rangle_{H^{\m}_{\sG}}$ decays exponentially to $0$ for $ \phi<\phi^{\crit}$, whereas it shows an algebraic decay in the regime  
$ \phi>\phi^{\crit}$ and incommensurate oscillations [see Fig.~\ref{fig:currPatterns_lowfill_corr}].

\section{Study of the large $\Jperp/J$-limit}
\label{sec:largeK}

% ------------------------------- FIGURE 4 ----------------------------------
\begin{figure}[t]
   \centering
    \resizebox{0.95\columnwidth}{!}{ \input{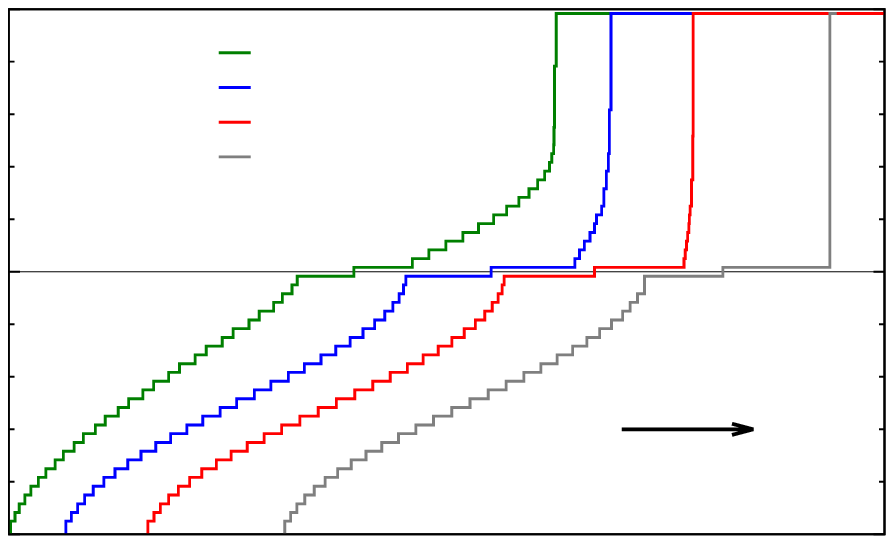} }
  \vspace{-0.01\textheight}
   \caption{(Color online) Equation of state $n=n(\mu)$ (density vs. chemical potential) of the effective spin-$\frac{1}{2}$ model Eq.~\eqref{suppmat:effectivespinhalf} for  $\phi=\pi$ and $\Jperp/J=2$, 4, 8 and $\infty$ (curves from \textit{left} to \textit{right}).
   System size is $L=60$ with open boundary conditions.
    The chemical potential $\mu$ has been rescaled and shifted for clarity.
}
   \label{fig:mag_eff}
\end{figure}
% ---------------------------------------------------------------------------

In the following we discuss the limit of strong rung tunneling $\Jperp/J\to \infty$ for the case of 
 hard-core bosons.

\subsection{Effective spin Hamiltonian}
In this regime and at filling $n \leq 0.5$, we may introduce a pseudo-spin-$\frac{1}{2}$ on a rung $r$ via
\begin{align} 
(\left|1,0 \right\rangle_{r} +e^{i r \phi} \left|0,1\right\rangle _{r})/\sqrt{2}\ &\to \left| \downarrow \right\rangle_{r} \nonumber \\
\left|0,0 \right\rangle_{r}  &\to \left| \uparrow \right\rangle_{r} .
\end{align}
Then the effective spin-$\frac{1}{2}$ model, to second order in $J/|\Jperp|$, contains two terms:
\begin{equation}
H_{1/2} = J\, H_{1/2}^0 + J^2/|\Jperp|\, H_{1/2}^1\, ,
\label{suppmat:effectivespinhalf}
\end{equation}
which are given by
\begin{align*}
H_{1/2}^0 &= \cos\left(\frac{\phi}{2}\right)\, \sum_{r} \left[ S^{\p}_{r} S^{\m}_{r+1} + {\rm h.c.} \right]\nonumber\\
H_{1/2}^1 &= -\cos\left(\frac{\phi}{2}\right)^2\, \sum_{r} \left[ S^{\p}_{r} (1/2 + S^z_{r+1}) S^{\m}_{r+2} + {\rm h.c.} \right] \nonumber\\
&\quad-\frac{1}{2}\sin\left(\frac{\phi}{2}\right)^2\, \sum_{r} \left[ S^{\p}_{r} (1/2 - S^z_{r+1}) S^{\m}_{r+2} + {\rm h.c.} \right] \nonumber\\
&\quad-\frac{1+3 \cos\left(\phi\right)}{2}\, \sum_{r} S^z_{r} S^z_{r+1}\,.
\end{align*}
In this effective model, zero magnetization corresponds to quarter filling $n=0.25$ in the original ladder model, while 
the fully polarized states correspond to zero or half filling. For small fluxes the first order term $H_{1/2}^0$
clearly dominates and the system behaves as a one-component Luttinger-liquid, and the central charge is thus $c=1$. 

\subsection{Behavior in the vicinity of  $\phi=\pi$}

To analyze the behavior of the system for finite fillings we use the effective model given in Eq.~\eqref{suppmat:effectivespinhalf}, which simplifies at $\phi=\pi$, and includes only a correlated next-nearest neighbor hopping term and nearest-neighbor 
Ising-type interactions:
\begin{align}
H_{1/2} = \frac{J^2}{2|\Jperp|} \sum_{r} \left( 2 S^z_{r} S^z_{r+1} - \Big[ S^{\p}_{r} (\frac{1}{2} - S^z_{r+1}) S^{\m}_{r+2} + 
{\rm h.c.} \Big] \right)\,.
\label{eq:effH12pi}
\end{align}
 Due to the correlated hopping in effective model Eq. (\ref{eq:effH12pi}), all tunneling processes are strongly suppressed and the Ising term proportional to $S^z_r S^z_{r+1}$ becomes dominant.
 Therefore, in the vicinity of $\phi=\pi$, the ground state at quarter filling $n=0.25$, expressed in terms of  effective spin degrees of freedom, is a doubly degenerate N\'eel state, which
 in the language of bosons translates into the 
charge density-wave state. In the $n=n(\mu)$ curves shown in Fig.~\ref{fig:mag_eff} this corresponds to the broad 
plateaux at $n=0.25$ for $\phi=\pi$ and different large values of $\Jperp/J$, which indicates a massive phase. 
When changing the flux from $\phi=\pi$, the system undergoes a transition from the 
N\'eel state ($\phi_{\cdw}^{\crit}<\phi\le \pi$) into a gapless XY phase ($\phi\le \phi_{\cdw}^{\crit}$) at $\phi=\phi_{\cdw}^{\crit}<\pi$.
The latter phase is characterized by a central charge of $c=1$.

The effective spin-$\frac{1}{2}$ model of Eq.~\eqref{eq:effH12pi}, obtained in the $\Jperp/J\to \infty$ limit, reveals another  interesting feature, namely 
metamagnetic behavior just below the saturation magnetization, which corresponds to a filling of $n=0.5$  for bosons as 
shown in Fig.~\ref{fig:mag_eff}. 
The magnetization curve exhibits a macroscopic jump to the saturation magnetization whose size increases with $\Jperp/J$. 
As one can see from Eq.~\eqref{eq:effH12pi}, for $\Jperp/J\to\infty$ and $n>0.25$, the ground-state energy 
is a linear function of $n$ and thus in the equation of state $n=n(\mu)$ 
(Fig.~\ref{fig:mag_eff}) the whole range of densities $0.25<n<0.5$ is unstable. 

\subsection{Perturbation theory at $n=0.5$}
\label{sec:largeK-pert}

As stated above (Sec.~\ref{sec:efftheory}), at half filling, \ie , for one boson per rung, the Hamiltonian can easily be diagonalized in the limit $\Jperp/J \to \infty$ we are considering.
The ground state is  a product of rung triplets in this phase for any flux $\phi$ and it is non-degenerate, separated by a finite gap from excited states.
In the $U/J \to \infty$ limit, there are two types of excitations.
The first type involves changing the total boson number by one by putting two bosons on the same rung  $\left|1,1\right\rangle_{r}$, or removing all bosons from one rung  $\left|0,0\right\rangle_{r}$, both at an energy cost of $\Jperp$.
The second type lives in the subspace of one boson per rung, and consists of exciting one rung of the ladder $r$ to a singlet
$(\left|1,0 \right>_{r} - e^{i r \phi} \left|0,1\right>_{r})/\sqrt{2}$, and is $2\Jperp$ higher in energy than the triplet.
Then, in perturbation theory and to first order in $J/\Jperp$, restoring a weak leg tunneling allows the bosons to hop to the neighbouring site, which favors local excitations of the type $\left|1,1\right>_{r} \otimes \left|0,0\right>_{r\pm 1}$.
One can then compute the expectation value of the chiral curent, which reads
\be
j_c = \frac{J^2}{2\Jperp} \sin(\phi)\,.
\label{eqsuppmat:largeK}
\ee
For finite but large $U \gg J$,  
besides the two types of excitations that we have discussed above for the case of $U/J \to \infty$,
one has to include 
double occupancies with an energy $\sim U$.
Retaining these three types of excited intermediate states, we calculate the ground-state energy in the lowest
(i.e. second) order in the intra-leg tunneling $J$ and obtain the chiral current
(as a derivative of the ground-state energy with the flux) as shown in Eq.~(6) of the main text.

\section{Phase diagram}

% ------------------------------- FIGURE 1 ----------------------------------
\begin{figure}[t]
   \centering
    \resizebox{0.95\columnwidth}{!}{ \input{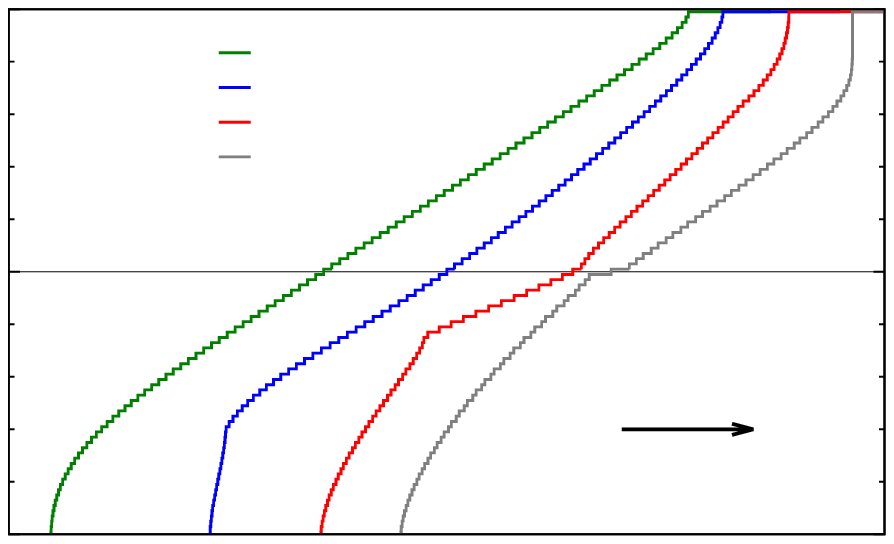} }
  \vspace{-0.01\textheight}
   \caption{(Color online) Equation of state $n=n(\mu)$ of the microscopic model Eq.~(1) of the main text, for $\Jperp/J=1.5$ and
   $\phi/\pi=0.4$, 0.6, 0.8 and 1 (curves from \textit{left} to \textit{right}). 
   System size is $L=100$ with open boundary conditions.
   The chemical potential $\mu$ has been shifted for clarity. }
   \label{fig:n_vs_mu}
\end{figure}
% ---------------------------------------------------------------------------
% ------------------------------- FIGURE 2 ----------------------------------
\begin{figure*}[t]
    \resizebox{1\textwidth}{!}{ \input{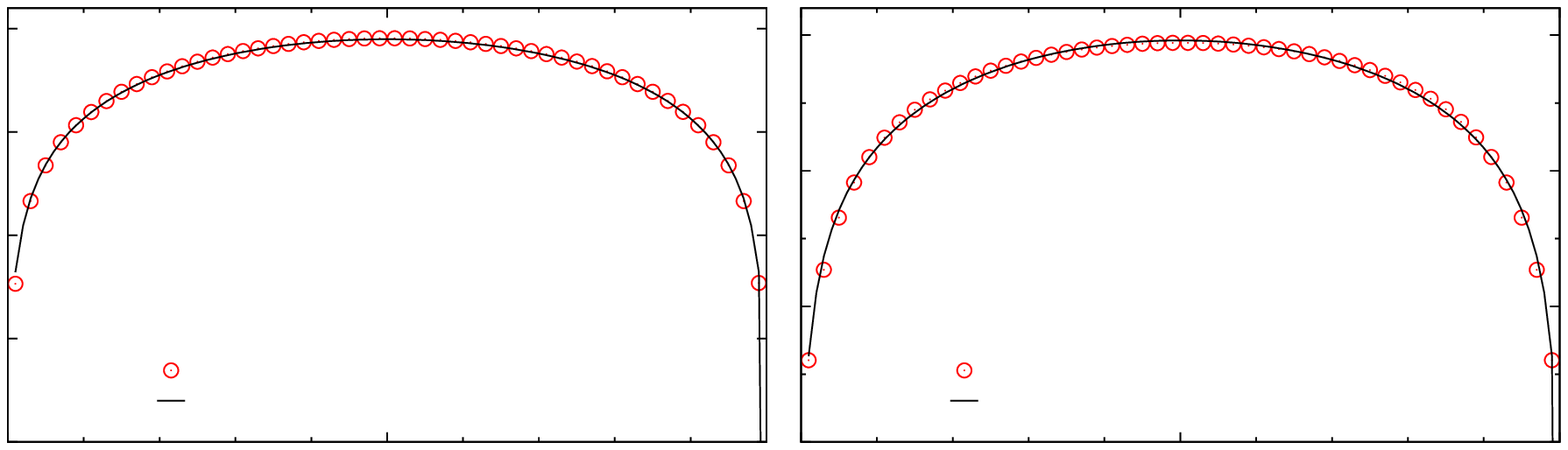} }
  \vspace{-0.01\textheight}
   \caption{(Color online) Von Neumann entropy $\SvN$ as a function of the partition size $\LA$.
   Example in (a) the M-SF phase ($\Jperp/J=1.5$, $n=0.25$ and $\phi=\pi/2$), 
   and (b)  the V-SF phase ($\Jperp/J=1.5$, $n=0.1$ and $\phi=0.9\pi$).
   The DMRG data (red dots) are obtained for $L=100$ and periodic boundary conditions on the legs.
   Solid black lines are fits of Eq.~\eqref{eq:entrop-fit} to the data; the fitting parameters $c$ and $g$ are given in the plots.}
   \label{fig:entrop_lowfill}
\end{figure*}
% ---------------------------------------------------------------------------
% ------------------------------- FIGURE 3 ----------------------------------
\begin{figure}[t]
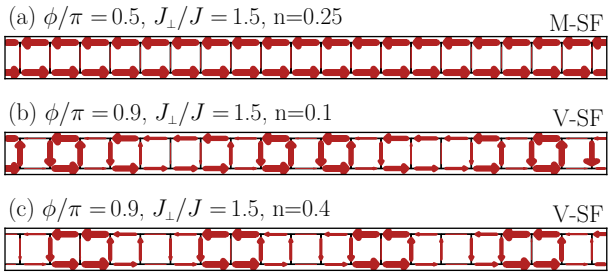

   \includegraphics[width=0.45\textwidth]{{{Beau-currents_pap_lowfill}}}\\
  \vspace{-0.01\textheight}
   \caption{(Color online) Current patterns corresponding to Fig.~\ref{fig:entrop_lowfill} (obtained with open boundary conditions).
  Examples in the M-SF phase  (a) for $\Jperp/J=1.5$, $n=0.25$ and $\phi=\pi/2$, 
   and the V-SF phase (b) for $\Jperp/J=1.5$, $n=0.1$ and $\phi=0.9\pi$ and (c) for $\Jperp/J=1.5$, $n=0.4$ and $\phi=0.9\pi$.
   In (c), part of the constant leg current has been filtered out to highlight the vortex pattern.
   }
      \label{fig:currPatterns_lowfill}
\end{figure}
% ---------------------------------------------------------------------------
% ------------------------------- FIGURE 3 ----------------------------------
\begin{figure}[t]
    \resizebox{0.45\textwidth}{!}{ \input{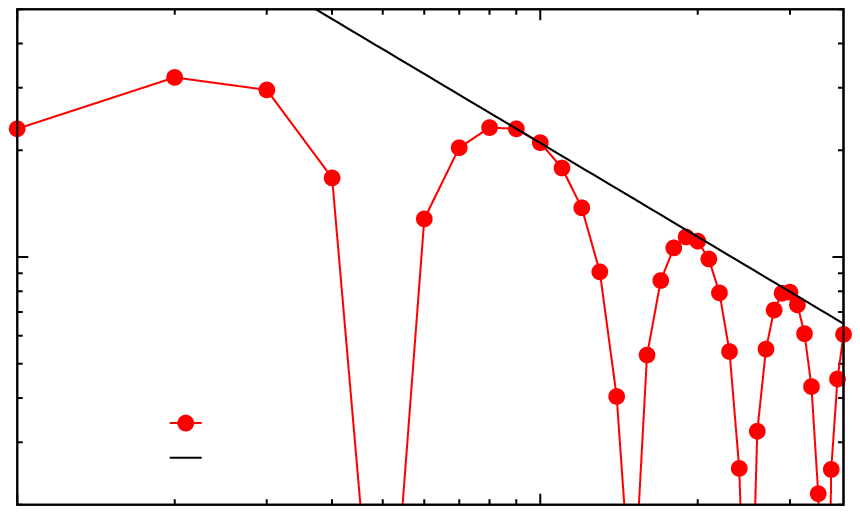} }\\
  \vspace{0.01\textheight}
    \resizebox{0.45\textwidth}{!}{ \input{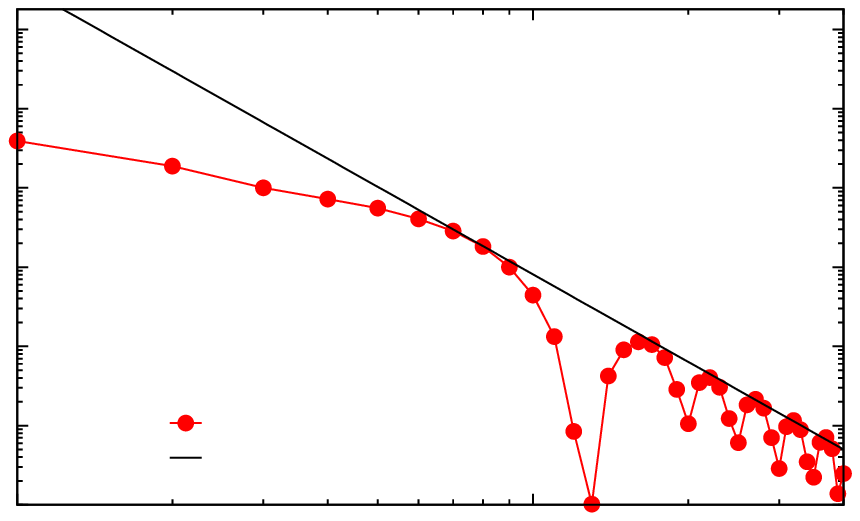} }
  \vspace{-0.01\textheight}
   \caption{(Color online) Rung-current correlation function $|\langle j^{\perpp}_r j^{\perpp}_{r'} \rangle |$ as a function of distance $|r-r'|$ in the V-SF phase (a) for $\Jperp/J=1.5$, $n=0.1$ and $\phi=0.9\pi$ and (b) for $\Jperp/J=1.5$, $n=0.4$ and $\phi=0.9\pi$. The solid line is a fit to the data using $|\langle j^{\perpp}_r j^{\perpp}_{r'} \rangle | \propto 1/|r-r'|^{\gamma}$.}
      \label{fig:currPatterns_lowfill_corr}
\end{figure}
% ---------------------------------------------------------------------------
In this section we present how the phase diagram, presented in Fig.~2 of the main text, has been obtained from DMRG simulations as a function of  $\Jperp/J$ and $\phi$.

\subsection{Equation of state}
Figure~\ref{fig:n_vs_mu} shows typical curves of the equation of state $n=n(\mu)$, where $\mu$ is the chemical potential, for $\Jperp/J=1.5$ and different values of $\phi$.
We see that some curves bear kinks, which are indicative of a phase transition between two gapless phases, with a change in the number of gapless excitations in Luttinger-liquid phases~\cite{okunishi03a}.
The curve for $\phi=0.4 \pi$ has no kink, whereas the curves at $\phi=0.6$ and $0.8\pi$ have one and two kinks, respectively. The positions of those kinks are reported in Fig.~2  of the main text and yield the phase boundaries.
Those curves also confirm, as stated in the main text, that the states at $n=0.5$ always have a charge gap.
At $\phi=\pi$, we find that the state at $n=0.25$ is gapped, and we see the metamagnetic transition to the state at $n=0.5$.
Both results are expected from the $\Jperp \gg J$ limit, as discussed in Sec.~\ref{sec:largeK}. 

\subsection{Von Neumann entropy \label{sec:vNentrop1}}
Further insight into the phase diagram comes from a careful analysis of the von Neumann entropy
\be
\SvN = - \mbox{tr}[\rhoA \mbox{ln}(\rhoA) ]\,
\ee
where $\rhoA$ is the reduced density matrix of a subsystem of length $\LA$ in a bipartition of the full system into two parts of linear size $\LA$ and $\LB$, with $L=\LA+\LB$ (in the bipartitioning, we cut the legs at the same point).
As an illustration, Fig.~\ref{fig:entrop_lowfill} presents $\SvN$, as a function of the partition size $\LA$ for two different points at $n \neq 0.5$.
Figure~\ref{fig:entrop_lowfill}(a) corresponds to a point in the small $\phi$ part of the phase diagram ($\Jperp/J=1.5$, $\phi=\pi/2$ and $n=0.25$), whereas Fig.~\ref{fig:entrop_lowfill}(b) is located in the phase found for high-fluxes ($\Jperp/J=1.5$, $\phi=0.9 \pi$ and $n=0.1$).
In conformal field theory, the von Neumann entropy is given by~\cite{Vidal03,Calabrese04}
\be
\SvN(\LA) = \frac{c}{3} \ln \left[ \frac{L}{\pi} \sin \left(\pi \frac{\LA}{L}\right) \right] + g,
\label{eq:entrop-fit}
\ee
in the case of a ladder geometry with periodic boundary conditions on the legs.
The central charge $c$ determines the number of gapless excitation modes in the system, whereas $g$ is a non-universal constant.
We performed fits of Eq.~\eqref{eq:entrop-fit} to the numerical data for the von Neumann entropy, which enabled us to measure $c$ and identify the different phases, thereby confirming the topology of the phase diagram.
We find that the phase at low $\phi$ has $c=1$ [see, e.g., Fig.~\ref{fig:entrop_lowfill}(a)], whereas we found $c=2$ at larger fluxes [see, e.g.,  Fig.~\ref{fig:entrop_lowfill}(b)].
This is consistent with the field-theoretical analysis presented in Sec.~\ref{sec:efftheory} at $\Jperp \ll J$, from which it follows that the antisymmetric mode, which is gapped at small flux $\phi$, becomes gapless at larger fluxes, while the symmetric mode is always gapless when $n \neq 0.5$.

\subsection{Current patterns}
Figures~\ref{fig:currPatterns_lowfill}(a) and (b) present sketches of the current patterns corresponding to the two parameter sets analyzed in Figs.~\ref{fig:entrop_lowfill}(a) and (b), respectively, as well as an additional one at $n=0.4$, i.e., in the upper lobe of the V-SF phase [Fig.~\ref{fig:entrop_lowfill}(c)].
We find that the $c=1$ phase [Fig.~\ref{fig:currPatterns_lowfill}(a)] has Meissner-like currents, i.e., a constant chiral current  $\jc$ along the legs and vanishing rung currents $\langle j^{\perpp}_r\rangle$.
In the $c=2$ phase, however, we find an oscillating leg current $\langle j^{\parallell}_{\ell,r}\rangle $  and non-vanishing rung currents $\langle j^{\perpp}_r\rangle $ [Fig.~\ref{fig:currPatterns_lowfill}(b) and (c)], which corresponds to vortex patterns.
This motivates us to call those phases Meissner-Superfluid (M-SF) and Vortex-Superfluid (V-SF), respectively.
Figures~\ref{fig:currPatterns_lowfill_corr}(a) and (b) show the rung-current correlation function in the V-SF phase, for those two last parameter sets.
We find that it indeed decreases algebraically with incommensurate oscillations, as predicted by field theory (see Sec.~\ref{sec:efftheory}).

\section{Mott Insulating phases at $n=0.5$}

% ------------------------------- FIGURE 5 ----------------------------------
\begin{figure*}[t]
  \centering
   \resizebox{1\textwidth}{!}{ \input{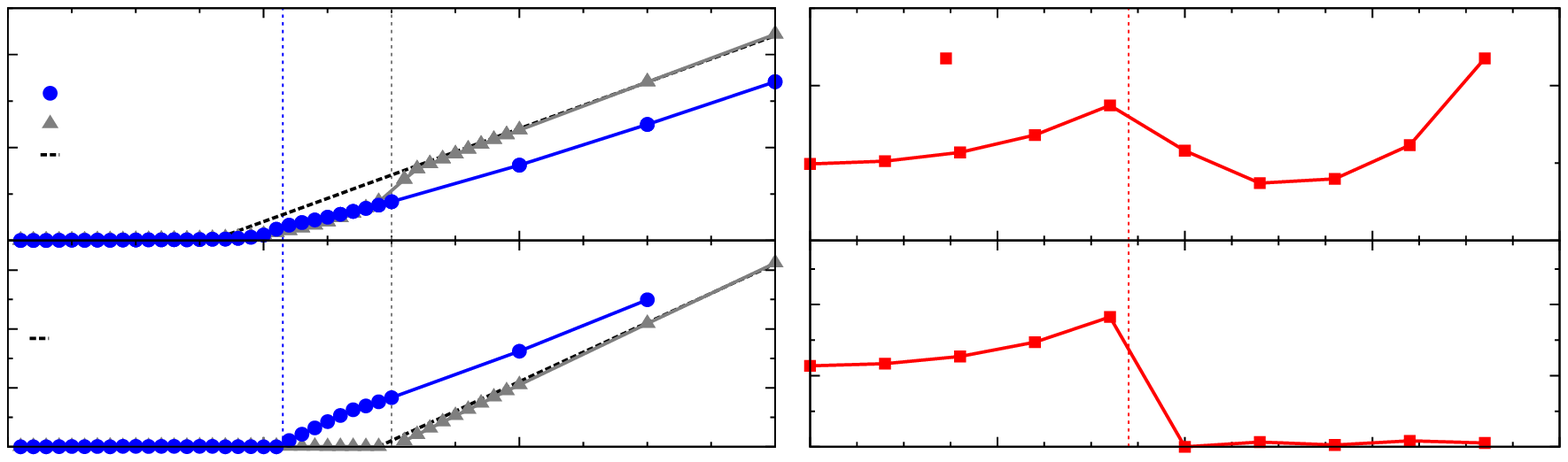} }
 \vspace{-0.01\textheight}
  \caption{(Color online) Gaps in the MI phases ($n=0.5$). 
  (a1)-(b1) Mass gap $\gapM$ and 
  (a2)-(b2) excitation gap $\gapEx$ in the subspace with constant $N$. Gaps are plotted as function of: (a1)-(a2) $\Jperp/J$  for $\phi/\pi=0.5$ and $\phi/\pi=0.8$, (b1)-(b2) flux $\phi$ for $\Jperp/J=1$. Dashed lines in (a1) and (a2) are fits of the $\phi=0.8 \pi$ cuts by $\gapM = \Jperp + A$ and $\gapEx = 2\Jperp + B$, respectively, with $A$ and $B$ as fitting parameters. Vertical dotted lines indicate the position of the V-MI to M-MI transition (estimated from Fig.~\ref{fig:curr_cuts_suppmat}). System size is $L=201$ with open boundary conditions.
  }
  \label{fig:gap_halffill}
\end{figure*}
% ---------------------------------------------------------------------------
% ------------------------------- FIGURE 6 ----------------------------------
\begin{figure*}[t]
   \centering
    \resizebox{1\textwidth}{!}{ \input{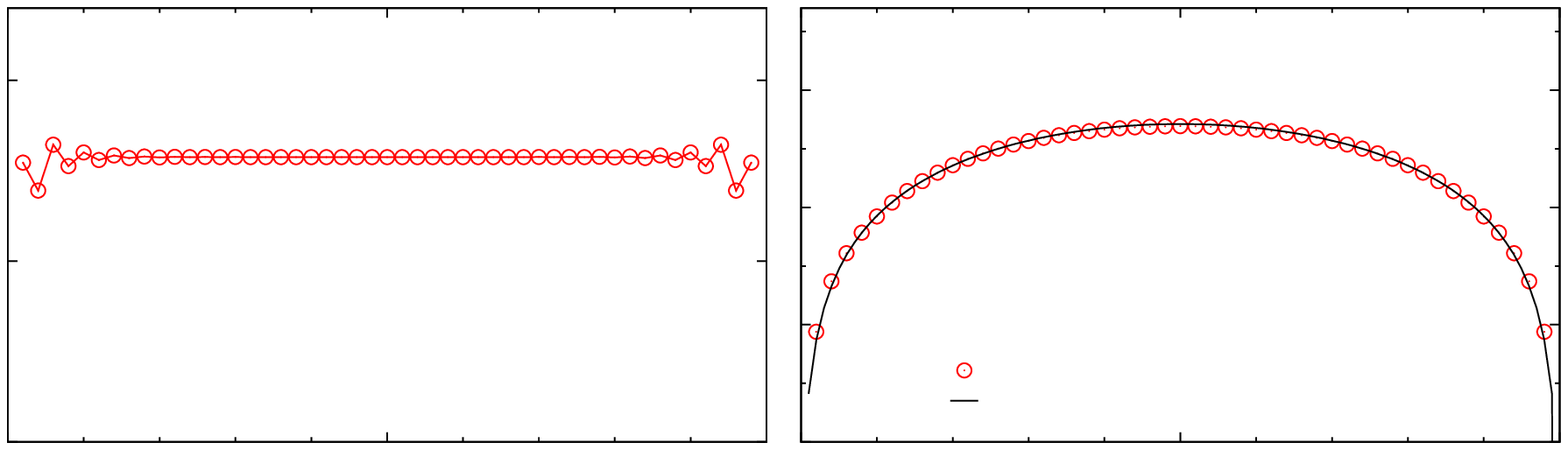} }\\
  \vspace{-0.01\textheight}
   \caption{(Color online) Von Neumann entropy $\SvN$ as a function of the partition size $\LA$.
    Example for (a) the M-MI phase ($\Jperp/J=1.5$, $n=0.5$ and $\phi=3\pi/4$), 
   and (b) the V-MI phase ($\Jperp/J=1.5$, $n=0.5$ and $\phi=0.8\pi$). The DMRG data are obtained for $L=50$
   and periodic boundary conditions on the legs.
   Solid black line on (b) is a fit of Eq.~\eqref{eq:entrop-fit} to the data; the fitting parameters $c$ and $g$ are indicated on the plots.
   }
   \label{fig:entrop_halffill}
\end{figure*}
% ---------------------------------------------------------------------------
% ------------------------------- FIGURE 7 ----------------------------------
\begin{figure*}[t]
   \centering
    \resizebox{1\textwidth}{!}{ \input{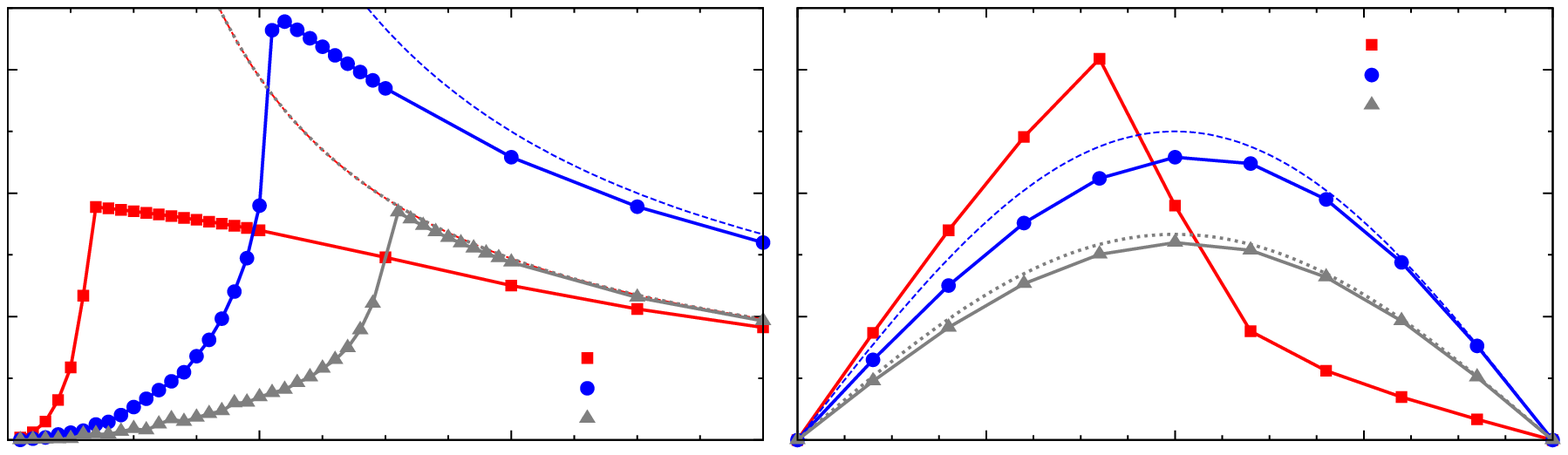} }
  \vspace{-0.01\textheight}
   \caption{(Color online) Cuts through the phase diagram of Fig.~3(a) of the main text.
   (a) Cuts at $\phi/\pi=0.2$, 0.5 and 0.8, and
   (b) cuts at $\Jperp/J=1$, 2 and 3
   The dashed lines are the theoretical predictions obtained for hard-core bosons when $\Jperp \gg J$ [see Eq.~\eqref{eqsuppmat:largeK}].
    System size is $L=201$ with open boundary conditions.
   }
   \label{fig:curr_cuts_suppmat}
\end{figure*}
% ---------------------------------------------------------------------------

In this section we present additional data for the Mott-insulating phases at $n=0.5$ with hard-core bosons, supporting the assertions of the main text.

\subsection{Excitation gaps}
Figure~\ref{fig:gap_halffill} shows results for  the mass gap
\be
\gapM = \frac{1}{2} \left[ E_{\gs}(N+1)+E_{\gs}(N-1) \right] -E_{\gs}(N) ,
\ee
where $E_{\gs}(N)$ is the energy of the ground state in the $N$-particles subspace, and for the excitation gap in the subspace with  fixed $N$,
\be
\gapEx = E_{\ex}(N)-E_{\gs}(N) ,
\ee
where $E_{\ex}(N)$ is the energy of the first excited state in the $N$ particles subspace, 
along cuts at $\phi=0.5\pi$ and $0.8\pi$ [Figs.~\ref{fig:gap_halffill}(a1) and (a2)] and along a cut at $\Jperp/J=1$ [Figs.~\ref{fig:gap_halffill}(b1) and (b2)].
Firstly, the key result is that the mass gap is finite, which numerically we are able to resolve for $J_{\perp}\gtrsim J $. This result applies  even to the V-MI phase [see, e.g., the data at $\phi=0.8 \pi$ around $\Jperp/J \sim 1.25$ in Fig.~\ref{fig:gap_halffill}(a1)], thus confirming that the system is a Mott insulator.
Field theory predicts an exponentially small mass gap $\gapM$ at $\Jperp \ll J$ (see Sec.~\ref{sec:efftheory} and Ref.~\cite{crepin11}).
This is hard to resolve numerically, yet the data plotted in Fig.~\ref{fig:gap_halffill}(a1) indeed suggest a rapid
decrease of the mass gap as $\Jperp$  goes to zero. In the limit of isolated rungs, $\Jperp \gg J$, with one boson per rung, one expects $\gapM \sim \Jperp$ (see Sec.~\ref{sec:largeK-pert}), which is consistent with the behaviour we find at large $\Jperp/J$, as highlighted by the fit of the $\phi=0.8\pi$ data by $\gapM=\Jperp+\textrm{cst}$.
Secondly, the excitation gap in the subspace with constant $N$, $\gapEx$, vanishes whenever the system is in the V-MI phase, \ie, at small $\Jperp/J$ in Fig.~\ref{fig:gap_halffill}(a2) and at high $\phi/\pi$ in Fig.~\ref{fig:gap_halffill}(b2).
In the $\Jperp \gg J$ limit, excitations that preserve $N$ have an energy $\gapEx \sim 2\Jperp$ (see Sec.~\ref{sec:largeK-pert}), which is consistent with the behaviour we find, as shown by the fit of $\gapM=2\Jperp+\textrm{cst}$ to the $\phi=0.8\pi$ data in Fig.~\ref{fig:gap_halffill}(a2).

\subsection{Von Neumann entropy}
We now complete our investigation by analyzing the von Neumann entropy of the $n=0.5$ phases.
We have carried out DMRG calculations of $\SvN$ as a function of the partition size $\LA$ (see Sec.~\ref{sec:vNentrop1}) in the M-MI and the V-MI phases, which are presented in Figs.~\ref{fig:entrop_halffill}(a) and (b) respectively.
In Fig.~\ref{fig:entrop_halffill}(a), obtained for $\Jperp/J=1.5$, $n=0.5$ and $\phi=3\pi/4$, the entropy saturates at a value $\SvN(L/2)$ independently of $\LA$, which is characteristic of an area law~\cite{schollwoeck11}.
In Fig.~\ref{fig:entrop_halffill}(b), in which $\Jperp/J=1.5$, $n=0.5$ and $\phi=0.8\pi$, we find the typical behaviour of a gapless system with $c=1$.
By fitting $\SvN(\LA)$ by Eq.~\eqref{eq:entrop-fit}, we obtain $c=1$.
Our findings are consistent with the $\Jperp \ll J$ field theory of Sec.~\ref{sec:efftheory}, which predicts that the symmetric mode is always gapped at half filling, whereas the antisymmetric mode is expected to be gapped at small flux, and can become gapless when $\phi$ increases.

\subsection{Chiral current}
In Fig.~\ref{fig:curr_cuts_suppmat} we provide more cuts through Fig.~3(b) of the main text, which shows the chiral current for hard-core bosons, thus complementing Fig.~4 of the main text.
Figure~\ref{fig:curr_cuts_suppmat}(a) shows  $\jc$ versus $\Jperp$ at $\phi=0.5\pi$ and $0.8\pi$.
The low $\Jperp/J$ behaviour is quadratic $\jc \sim \Jperp^2$, with a prefactor that decreases with increasing flux, which is thus compatible with the prediction~Eq.~\eqref{chircur} for small fluxes (as one has $\Km=1$ in the hard-core limit).
The large $\Jperp/J$ behaviour is very well captured by standard perturbation theory [see Eq.~(6) of the main text], with the asymptotic behaviour $\jc \sim 1/\Jperp$, as shown by the dashed lines.
Finally, one can notice that the transition from the V-MI to the M-MI is always associated to a kink in the chiral current.
Figure~\ref{fig:curr_cuts_suppmat}(b) shows cuts of $\jc$ at $\Jperp=J$, $2J$ and $3J$.
The $\jc(\phi) \sim \sin(\phi)$ behaviour predicted in the $\Jperp \gg J$ limit [Eq.~\eqref{eqsuppmat:largeK}] is generic in the M-MI phase, and the transition to the V-MI phase is once again associated 
with a kink at $\phi^{\rm cr}$.

% ------------------------------- FIGURE 8 ----------------------------------
\begin{figure}[t]
    \resizebox{0.95\columnwidth}{!}{ \input{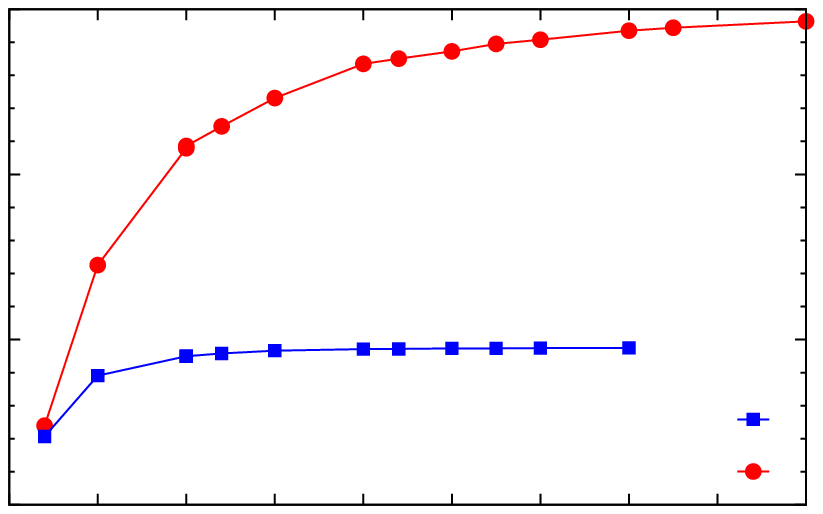} }
  \vspace{-0.01\textheight}
   \caption{(Color online) Convergence of the von Neumann entropy for the central bipartition $\SvN(L/2)$ as a function of the dimension of the matrix product state $m$, for the parameters of Fig.~\ref{fig:entrop_lowfill}, in the M-SF phase, where $c=1$, (squares) and in the V-SF phase, where $c=2$ (dots).}
      \label{fig:conv_entrop}
\end{figure}
% ---------------------------------------------------------------------------

\section{Details on DMRG data}
Let us finally give a few details on the numerical quality of the DMRG data that are provided in this work, where we use a finite-size DMRG algorithm \cite{schollwoeck05}.
In the calculations of currents and the equation of state $n(\mu)$ we use systems with up to $L=201$ rungs and  open boundary conditions.
For $U/J < \infty$, the local basis is restricted to at maximum  four bosons per site, and we have checked that calculations with three bosons yield consistent results.
 Energies are typically converged up to the $8^{\rm th}$ digit.

Periodic boundary conditions are used for the calculation of the von Neumann entropy shown in Figs.~\ref{fig:entrop_lowfill} and \ref{fig:entrop_halffill}. 
Figure~\ref{fig:conv_entrop} illustrates the convergence of  $\SvN(L/2)$ as a function of the dimension of the matrix product state $m$ for the two parameter sets of Fig.~\ref{fig:entrop_lowfill}.
In general, we keep at least $m=1000$ states in the matrix product state representation, typically $m=2500$ [\eg\ in Fig.~\ref{fig:entrop_lowfill}(a)] and up to $m = 4500$ where necessary [\eg\ in Fig.~\ref{fig:entrop_lowfill}(b)].
The $c=2$ states are the hardest to converge, the extraction of the central charge as $c=2$ is, however, robust, 
which we also verified by doing runs for open boundary conditions.


\begin{thebibliography}{59}%
\makeatletter
\providecommand \@ifxundefined [1]{%
 \@ifx{#1\undefined}
}%
\providecommand \@ifnum [1]{%
 \ifnum #1\expandafter \@firstoftwo
 \else \expandafter \@secondoftwo
 \fi
}%
\providecommand \@ifx [1]{%
 \ifx #1\expandafter \@firstoftwo
 \else \expandafter \@secondoftwo
 \fi
}%
\providecommand \natexlab [1]{#1}%
\providecommand \enquote  [1]{``#1''}%
\providecommand \bibnamefont  [1]{#1}%
\providecommand \bibfnamefont [1]{#1}%
\providecommand \citenamefont [1]{#1}%
\providecommand \href@noop [0]{\@secondoftwo}%
\providecommand \href [0]{\begingroup \@sanitize@url \@href}%
\providecommand \@href[1]{\@@startlink{#1}\@@href}%
\providecommand \@@href[1]{\endgroup#1\@@endlink}%
\providecommand \@sanitize@url [0]{\catcode `\\12\catcode `\$12\catcode
  `\&12\catcode `\#12\catcode `\^12\catcode `\_12\catcode `\%12\relax}%
\providecommand \@@startlink[1]{}%
\providecommand \@@endlink[0]{}%
\providecommand \url  [0]{\begingroup\@sanitize@url \@url }%
\providecommand \@url [1]{\endgroup\@href {#1}{\urlprefix }}%
\providecommand \urlprefix  [0]{URL }%
\providecommand \Eprint [0]{\href }%
\providecommand \doibase [0]{http://dx.doi.org/}%
\providecommand \selectlanguage [0]{\@gobble}%
\providecommand \bibinfo  [0]{\@secondoftwo}%
\providecommand \bibfield  [0]{\@secondoftwo}%
\providecommand \translation [1]{[#1]}%
\providecommand \BibitemOpen [0]{}%
\providecommand \bibitemStop [0]{}%
\providecommand \bibitemNoStop [0]{.\EOS\space}%
\providecommand \EOS [0]{\spacefactor3000\relax}%
\providecommand \BibitemShut  [1]{\csname bibitem#1\endcsname}%
\let\auto@bib@innerbib\@empty
%</preamble>
\bibitem [{\citenamefont {Thouless}\ \emph {et~al.}(1982)\citenamefont
  {Thouless}, \citenamefont {Kohmoto}, \citenamefont {Nightingale},\ and\
  \citenamefont {den Nijs}}]{Thouless82}%
  \BibitemOpen
  \bibfield  {author} {\bibinfo {author} {\bibfnamefont {D.~J.}\ \bibnamefont
  {Thouless}}, \bibinfo {author} {\bibfnamefont {M.}~\bibnamefont {Kohmoto}},
  \bibinfo {author} {\bibfnamefont {M.~P.}\ \bibnamefont {Nightingale}}, \ and\
  \bibinfo {author} {\bibfnamefont {M.}~\bibnamefont {den Nijs}},\ }\href@noop
  {} {\bibfield  {journal} {\bibinfo  {journal} {Phys. Rev. Lett.}\ }\textbf
  {\bibinfo {volume} {49}},\ \bibinfo {pages} {405} (\bibinfo {year}
  {1982})}\BibitemShut {NoStop}%
\bibitem [{\citenamefont {Kane}\ and\ \citenamefont {Mele}(2005)}]{kane05}%
  \BibitemOpen
  \bibfield  {author} {\bibinfo {author} {\bibfnamefont {C.~L.}\ \bibnamefont
  {Kane}}\ and\ \bibinfo {author} {\bibfnamefont {E.~J.}\ \bibnamefont
  {Mele}},\ }\href@noop {} {\bibfield  {journal} {\bibinfo  {journal} {Phys.
  Rev. Lett.}\ }\textbf {\bibinfo {volume} {95}},\ \bibinfo {pages} {146802}
  (\bibinfo {year} {2005})}\BibitemShut {NoStop}%
\bibitem [{\citenamefont {Hasan}\ and\ \citenamefont {Kane}(2010)}]{hasan10}%
  \BibitemOpen
  \bibfield  {author} {\bibinfo {author} {\bibfnamefont {M.~Z.}\ \bibnamefont
  {Hasan}}\ and\ \bibinfo {author} {\bibfnamefont {C.~L.}\ \bibnamefont
  {Kane}},\ }\href@noop {} {\bibfield  {journal} {\bibinfo  {journal} {Rev.
  Mod. Phys.}\ }\textbf {\bibinfo {volume} {82}},\ \bibinfo {pages} {3045}
  (\bibinfo {year} {2010})}\BibitemShut {NoStop}%
\bibitem [{\citenamefont {Qi}\ and\ \citenamefont {Zhang}(2011)}]{qi11}%
  \BibitemOpen
  \bibfield  {author} {\bibinfo {author} {\bibfnamefont {X.-L.}\ \bibnamefont
  {Qi}}\ and\ \bibinfo {author} {\bibfnamefont {S.-C.}\ \bibnamefont {Zhang}},\
  }\href@noop {} {\bibfield  {journal} {\bibinfo  {journal} {Rev. Mod. Phys.}\
  }\textbf {\bibinfo {volume} {83}},\ \bibinfo {pages} {1057} (\bibinfo {year}
  {2011})}\BibitemShut {NoStop}%
\bibitem [{\citenamefont {Kitaev}(2001)}]{kitaev00}%
  \BibitemOpen
  \bibfield  {author} {\bibinfo {author} {\bibfnamefont {A.}~\bibnamefont
  {Kitaev}},\ }\href@noop {} {\bibfield  {journal} {\bibinfo  {journal}
  {Phys.-Usp.}\ }\textbf {\bibinfo {volume} {44}},\ \bibinfo {pages} {131}
  (\bibinfo {year} {2001})}\BibitemShut {NoStop}%
\bibitem [{\citenamefont {Fu}\ and\ \citenamefont {Kane}(2008)}]{Fu08}%
  \BibitemOpen
  \bibfield  {author} {\bibinfo {author} {\bibfnamefont {L.}~\bibnamefont
  {Fu}}\ and\ \bibinfo {author} {\bibfnamefont {C.~L.}\ \bibnamefont {Kane}},\
  }\href@noop {} {\bibfield  {journal} {\bibinfo  {journal} {Phys. Rev. Lett.}\
  }\textbf {\bibinfo {volume} {100}},\ \bibinfo {pages} {096407} (\bibinfo
  {year} {2008})}\BibitemShut {NoStop}%
\bibitem [{\citenamefont {Dalibard}\ \emph {et~al.}(2011)\citenamefont
  {Dalibard}, \citenamefont {Gerbier}, \citenamefont
  {Juzeli\ifmmode~\bar{u}\else \={u}\fi{}nas},\ and\ \citenamefont
  {\"Ohberg}}]{dalibard11}%
  \BibitemOpen
  \bibfield  {author} {\bibinfo {author} {\bibfnamefont {J.}~\bibnamefont
  {Dalibard}}, \bibinfo {author} {\bibfnamefont {F.}~\bibnamefont {Gerbier}},
  \bibinfo {author} {\bibfnamefont {G.}~\bibnamefont
  {Juzeli\ifmmode~\bar{u}\else \={u}\fi{}nas}}, \ and\ \bibinfo {author}
  {\bibfnamefont {P.}~\bibnamefont {\"Ohberg}},\ }\href {\doibase
  10.1103/RevModPhys.83.1523} {\bibfield  {journal} {\bibinfo  {journal} {Rev.
  Mod. Phys.}\ }\textbf {\bibinfo {volume} {83}},\ \bibinfo {pages} {1523}
  (\bibinfo {year} {2011})}\BibitemShut {NoStop}%
\bibitem [{\citenamefont {Lin}\ \emph {et~al.}(2009)\citenamefont {Lin},
  \citenamefont {Compton}, \citenamefont {Jim\'enez-Garc\'{\i}a}, \citenamefont
  {Porto},\ and\ \citenamefont {Spielman}}]{lin09}%
  \BibitemOpen
  \bibfield  {author} {\bibinfo {author} {\bibfnamefont {Y.-J.}\ \bibnamefont
  {Lin}}, \bibinfo {author} {\bibfnamefont {R.~L.}\ \bibnamefont {Compton}},
  \bibinfo {author} {\bibfnamefont {K.}~\bibnamefont {Jim\'enez-Garc\'{\i}a}},
  \bibinfo {author} {\bibfnamefont {J.~V.}\ \bibnamefont {Porto}}, \ and\
  \bibinfo {author} {\bibfnamefont {I.~B.}\ \bibnamefont {Spielman}},\
  }\href@noop {} {\bibfield  {journal} {\bibinfo  {journal} {Nature (London)}\
  }\textbf {\bibinfo {volume} {462}},\ \bibinfo {pages} {628} (\bibinfo {year}
  {2009})}\BibitemShut {NoStop}%
\bibitem [{\citenamefont {Lin}\ \emph {et~al.}(2011)\citenamefont {Lin},
  \citenamefont {Compton}, \citenamefont {Jim\'enez-Garc\'{\i}a}, \citenamefont
  {Phillips}, \citenamefont {Porto},\ and\ \citenamefont {Spielman}}]{lin11}%
  \BibitemOpen
  \bibfield  {author} {\bibinfo {author} {\bibfnamefont {Y.-J.}\ \bibnamefont
  {Lin}}, \bibinfo {author} {\bibfnamefont {R.~L.}\ \bibnamefont {Compton}},
  \bibinfo {author} {\bibnamefont {Jim\'enez-Garc\'{\i}a}}, \bibinfo {author}
  {\bibfnamefont {W.~D.}\ \bibnamefont {Phillips}}, \bibinfo {author}
  {\bibfnamefont {J.~V.}\ \bibnamefont {Porto}}, \ and\ \bibinfo {author}
  {\bibfnamefont {I.~B.}\ \bibnamefont {Spielman}},\ }\href@noop {} {\bibfield
  {journal} {\bibinfo  {journal} {Nature Phys.}\ }\textbf {\bibinfo {volume}
  {7}},\ \bibinfo {pages} {531} (\bibinfo {year} {2011})}\BibitemShut {NoStop}%
\bibitem [{\citenamefont {Jim\'enez-Garc\'{\i}a}\ \emph
  {et~al.}(2012)\citenamefont {Jim\'enez-Garc\'{\i}a}, \citenamefont {LeBlanc},
  \citenamefont {Williams}, \citenamefont {Beeler}, \citenamefont {Perry},\
  and\ \citenamefont {Spielman}}]{jimenez12}%
  \BibitemOpen
  \bibfield  {author} {\bibinfo {author} {\bibfnamefont {K.}~\bibnamefont
  {Jim\'enez-Garc\'{\i}a}}, \bibinfo {author} {\bibfnamefont {L.~J.}\
  \bibnamefont {LeBlanc}}, \bibinfo {author} {\bibfnamefont {R.~A.}\
  \bibnamefont {Williams}}, \bibinfo {author} {\bibfnamefont {M.~C.}\
  \bibnamefont {Beeler}}, \bibinfo {author} {\bibfnamefont {A.~R.}\
  \bibnamefont {Perry}}, \ and\ \bibinfo {author} {\bibfnamefont {I.~B.}\
  \bibnamefont {Spielman}},\ }\href@noop {} {\bibfield  {journal} {\bibinfo
  {journal} {Phys. Rev. Lett.}\ }\textbf {\bibinfo {volume} {108}},\ \bibinfo
  {pages} {225303} (\bibinfo {year} {2012})}\BibitemShut {NoStop}%
\bibitem [{\citenamefont {Aidelsburger}\ \emph {et~al.}(2011)\citenamefont
  {Aidelsburger}, \citenamefont {Atala}, \citenamefont {Nascimb\`ene},
  \citenamefont {Trotzky}, \citenamefont {Chen},\ and\ \citenamefont
  {Bloch}}]{aidelsburger11}%
  \BibitemOpen
  \bibfield  {author} {\bibinfo {author} {\bibfnamefont {M.}~\bibnamefont
  {Aidelsburger}}, \bibinfo {author} {\bibfnamefont {M.}~\bibnamefont {Atala}},
  \bibinfo {author} {\bibfnamefont {S.}~\bibnamefont {Nascimb\`ene}}, \bibinfo
  {author} {\bibfnamefont {S.}~\bibnamefont {Trotzky}}, \bibinfo {author}
  {\bibfnamefont {Y.-A.}\ \bibnamefont {Chen}}, \ and\ \bibinfo {author}
  {\bibfnamefont {I.}~\bibnamefont {Bloch}},\ }\href@noop {} {\bibfield
  {journal} {\bibinfo  {journal} {Phys. Rev. Lett.}\ }\textbf {\bibinfo
  {volume} {107}},\ \bibinfo {pages} {255301} (\bibinfo {year}
  {2011})}\BibitemShut {NoStop}%
\bibitem [{\citenamefont {Struck}\ \emph {et~al.}(2012)\citenamefont {Struck},
  \citenamefont {\"Olschl\"ager}, \citenamefont {Weinberg}, \citenamefont
  {Hauke}, \citenamefont {Simonet}, \citenamefont {Eckardt}, \citenamefont
  {Lewenstein}, \citenamefont {Sengstock},\ and\ \citenamefont
  {Windpassinger}}]{struck12}%
  \BibitemOpen
  \bibfield  {author} {\bibinfo {author} {\bibfnamefont {J.}~\bibnamefont
  {Struck}}, \bibinfo {author} {\bibfnamefont {C.}~\bibnamefont
  {\"Olschl\"ager}}, \bibinfo {author} {\bibfnamefont {M.}~\bibnamefont
  {Weinberg}}, \bibinfo {author} {\bibfnamefont {P.}~\bibnamefont {Hauke}},
  \bibinfo {author} {\bibfnamefont {J.}~\bibnamefont {Simonet}}, \bibinfo
  {author} {\bibfnamefont {A.}~\bibnamefont {Eckardt}}, \bibinfo {author}
  {\bibfnamefont {M.}~\bibnamefont {Lewenstein}}, \bibinfo {author}
  {\bibfnamefont {K.}~\bibnamefont {Sengstock}}, \ and\ \bibinfo {author}
  {\bibfnamefont {P.}~\bibnamefont {Windpassinger}},\ }\href@noop {} {\bibfield
   {journal} {\bibinfo  {journal} {Phys. Rev. Lett.}\ }\textbf {\bibinfo
  {volume} {108}},\ \bibinfo {pages} {225304} (\bibinfo {year}
  {2012})}\BibitemShut {NoStop}%
\bibitem [{\citenamefont {Aidelsburger}\ \emph {et~al.}(2013)\citenamefont
  {Aidelsburger}, \citenamefont {Atala}, \citenamefont {Lohse}, \citenamefont
  {Barreiro}, \citenamefont {Paredes},\ and\ \citenamefont
  {Bloch}}]{aidelsburger13}%
  \BibitemOpen
  \bibfield  {author} {\bibinfo {author} {\bibfnamefont {M.}~\bibnamefont
  {Aidelsburger}}, \bibinfo {author} {\bibfnamefont {M.}~\bibnamefont {Atala}},
  \bibinfo {author} {\bibfnamefont {M.}~\bibnamefont {Lohse}}, \bibinfo
  {author} {\bibfnamefont {J.~T.}\ \bibnamefont {Barreiro}}, \bibinfo {author}
  {\bibfnamefont {B.}~\bibnamefont {Paredes}}, \ and\ \bibinfo {author}
  {\bibfnamefont {I.}~\bibnamefont {Bloch}},\ }\href@noop {} {\bibfield
  {journal} {\bibinfo  {journal} {Phys. Rev. Lett.}\ }\textbf {\bibinfo
  {volume} {111}},\ \bibinfo {pages} {185301} (\bibinfo {year}
  {2013})}\BibitemShut {NoStop}%
\bibitem [{\citenamefont {Miyake}\ \emph {et~al.}(2013)\citenamefont {Miyake},
  \citenamefont {Siviloglou}, \citenamefont {Kennedy}, \citenamefont {Burton},\
  and\ \citenamefont {Ketterle}}]{miyake13}%
  \BibitemOpen
  \bibfield  {author} {\bibinfo {author} {\bibfnamefont {H.}~\bibnamefont
  {Miyake}}, \bibinfo {author} {\bibfnamefont {G.~A.}\ \bibnamefont
  {Siviloglou}}, \bibinfo {author} {\bibfnamefont {C.~J.}\ \bibnamefont
  {Kennedy}}, \bibinfo {author} {\bibfnamefont {W.~C.}\ \bibnamefont {Burton}},
  \ and\ \bibinfo {author} {\bibfnamefont {W.}~\bibnamefont {Ketterle}},\
  }\href@noop {} {\bibfield  {journal} {\bibinfo  {journal} {Phys. Rev. Lett.}\
  }\textbf {\bibinfo {volume} {111}},\ \bibinfo {pages} {185302} (\bibinfo
  {year} {2013})}\BibitemShut {NoStop}%
\bibitem [{\citenamefont {S\o{}rensen}\ \emph {et~al.}(2005)\citenamefont
  {S\o{}rensen}, \citenamefont {Demler},\ and\ \citenamefont
  {Lukin}}]{sorensen05}%
  \BibitemOpen
  \bibfield  {author} {\bibinfo {author} {\bibfnamefont {A.}~\bibnamefont
  {S\o{}rensen}}, \bibinfo {author} {\bibfnamefont {E.}~\bibnamefont {Demler}},
  \ and\ \bibinfo {author} {\bibfnamefont {M.}~\bibnamefont {Lukin}},\ }\href
  {\doibase 10.1103/PhysRevLett.94.086803} {\bibfield  {journal} {\bibinfo
  {journal} {Phys. Rev. Lett.}\ }\textbf {\bibinfo {volume} {94}},\ \bibinfo
  {pages} {086803} (\bibinfo {year} {2005})}\BibitemShut {NoStop}%
\bibitem [{\citenamefont {Palmer}\ and\ \citenamefont
  {Jaksch}(2006)}]{palmer06}%
  \BibitemOpen
  \bibfield  {author} {\bibinfo {author} {\bibfnamefont {R.}~\bibnamefont
  {Palmer}}\ and\ \bibinfo {author} {\bibfnamefont {D.}~\bibnamefont
  {Jaksch}},\ }\href {\doibase 10.1103/PhysRevLett.96.180407} {\bibfield
  {journal} {\bibinfo  {journal} {Phys. Rev. Lett.}\ }\textbf {\bibinfo
  {volume} {96}},\ \bibinfo {pages} {180407} (\bibinfo {year}
  {2006})}\BibitemShut {NoStop}%
\bibitem [{\citenamefont {Hafezi}\ \emph {et~al.}(2007)\citenamefont {Hafezi},
  \citenamefont {S\o{}rensen}, \citenamefont {Demler},\ and\ \citenamefont
  {Lukin}}]{hafezi07}%
  \BibitemOpen
  \bibfield  {author} {\bibinfo {author} {\bibfnamefont {M.}~\bibnamefont
  {Hafezi}}, \bibinfo {author} {\bibfnamefont {A.}~\bibnamefont {S\o{}rensen}},
  \bibinfo {author} {\bibfnamefont {E.}~\bibnamefont {Demler}}, \ and\ \bibinfo
  {author} {\bibfnamefont {M.}~\bibnamefont {Lukin}},\ }\href {\doibase
  10.1103/PhysRevA.76.023613} {\bibfield  {journal} {\bibinfo  {journal} {Phys.
  Rev. A}\ }\textbf {\bibinfo {volume} {76}},\ \bibinfo {pages} {023613}
  (\bibinfo {year} {2007})}\BibitemShut {NoStop}%
\bibitem [{\citenamefont {Cooper}(2008)}]{cooper08}%
  \BibitemOpen
  \bibfield  {author} {\bibinfo {author} {\bibfnamefont {N.~R.}\ \bibnamefont
  {Cooper}},\ }\href@noop {} {\bibfield  {journal} {\bibinfo  {journal}
  {Advances in Physics}\ }\textbf {\bibinfo {volume} {57}},\ \bibinfo {pages}
  {539} (\bibinfo {year} {2008})}\BibitemShut {NoStop}%
\bibitem [{\citenamefont {Fetter}(2009)}]{fetter09}%
  \BibitemOpen
  \bibfield  {author} {\bibinfo {author} {\bibfnamefont {A.}~\bibnamefont
  {Fetter}},\ }\href {\doibase 10.1103/RevModPhys.81.647} {\bibfield  {journal}
  {\bibinfo  {journal} {Rev. Mod. Phys.}\ }\textbf {\bibinfo {volume} {81}},\
  \bibinfo {pages} {647} (\bibinfo {year} {2009})}\BibitemShut {NoStop}%
\bibitem [{\citenamefont {M\"oller}\ and\ \citenamefont
  {Cooper}(2009)}]{moeller09}%
  \BibitemOpen
  \bibfield  {author} {\bibinfo {author} {\bibfnamefont {G.}~\bibnamefont
  {M\"oller}}\ and\ \bibinfo {author} {\bibfnamefont {N.}~\bibnamefont
  {Cooper}},\ }\href {\doibase 10.1103/PhysRevLett.103.105303} {\bibfield
  {journal} {\bibinfo  {journal} {Phys. Rev. Lett.}\ }\textbf {\bibinfo
  {volume} {103}},\ \bibinfo {pages} {105303} (\bibinfo {year}
  {2009})}\BibitemShut {NoStop}%
\bibitem [{\citenamefont {Senthil}\ and\ \citenamefont
  {Levin}(2013)}]{senthil13}%
  \BibitemOpen
  \bibfield  {author} {\bibinfo {author} {\bibfnamefont {T.}~\bibnamefont
  {Senthil}}\ and\ \bibinfo {author} {\bibfnamefont {M.}~\bibnamefont
  {Levin}},\ }\href@noop {} {\bibfield  {journal} {\bibinfo  {journal} {Phys.
  Rev. Lett.}\ }\textbf {\bibinfo {volume} {110}},\ \bibinfo {pages} {046801}
  (\bibinfo {year} {2013})}\BibitemShut {NoStop}%
\bibitem [{\citenamefont {Regnault}\ and\ \citenamefont
  {Senthil}(2013)}]{regnault13}%
  \BibitemOpen
  \bibfield  {author} {\bibinfo {author} {\bibfnamefont {N.}~\bibnamefont
  {Regnault}}\ and\ \bibinfo {author} {\bibfnamefont {T.}~\bibnamefont
  {Senthil}},\ }\href@noop {} {\bibfield  {journal} {\bibinfo  {journal} {Phys.
  Rev. B}\ }\textbf {\bibinfo {volume} {88}},\ \bibinfo {pages} {161106}
  (\bibinfo {year} {2013})}\BibitemShut {NoStop}%
\bibitem [{\citenamefont {Cole}\ \emph {et~al.}(2012)\citenamefont {Cole},
  \citenamefont {Zhang}, \citenamefont {Paramekanti},\ and\ \citenamefont
  {Trivedi}}]{cole12}%
  \BibitemOpen
  \bibfield  {author} {\bibinfo {author} {\bibfnamefont {W.~S.}\ \bibnamefont
  {Cole}}, \bibinfo {author} {\bibfnamefont {S.}~\bibnamefont {Zhang}},
  \bibinfo {author} {\bibfnamefont {A.}~\bibnamefont {Paramekanti}}, \ and\
  \bibinfo {author} {\bibfnamefont {N.}~\bibnamefont {Trivedi}},\ }\href@noop
  {} {\bibfield  {journal} {\bibinfo  {journal} {Phys. Rev. Lett.}\ }\textbf
  {\bibinfo {volume} {109}},\ \bibinfo {pages} {085302} (\bibinfo {year}
  {2012})}\BibitemShut {NoStop}%
\bibitem [{\citenamefont {Radi\ifmmode~\acute{c}\else \'{c}\fi{}}\ \emph
  {et~al.}(2012)\citenamefont {Radi\ifmmode~\acute{c}\else \'{c}\fi{}},
  \citenamefont {Di~Ciolo}, \citenamefont {Sun},\ and\ \citenamefont
  {Galitski}}]{radic12}%
  \BibitemOpen
  \bibfield  {author} {\bibinfo {author} {\bibfnamefont {J.}~\bibnamefont
  {Radi\ifmmode~\acute{c}\else \'{c}\fi{}}}, \bibinfo {author} {\bibfnamefont
  {A.}~\bibnamefont {Di~Ciolo}}, \bibinfo {author} {\bibfnamefont
  {K.}~\bibnamefont {Sun}}, \ and\ \bibinfo {author} {\bibfnamefont
  {V.}~\bibnamefont {Galitski}},\ }\href@noop {} {\bibfield  {journal}
  {\bibinfo  {journal} {Phys. Rev. Lett.}\ }\textbf {\bibinfo {volume} {109}},\
  \bibinfo {pages} {085303} (\bibinfo {year} {2012})}\BibitemShut {NoStop}%
\bibitem [{\citenamefont {Cai}\ \emph {et~al.}(2012)\citenamefont {Cai},
  \citenamefont {Zhou},\ and\ \citenamefont {Wu}}]{cai12}%
  \BibitemOpen
  \bibfield  {author} {\bibinfo {author} {\bibfnamefont {Z.}~\bibnamefont
  {Cai}}, \bibinfo {author} {\bibfnamefont {X.}~\bibnamefont {Zhou}}, \ and\
  \bibinfo {author} {\bibfnamefont {C.}~\bibnamefont {Wu}},\ }\href@noop {}
  {\bibfield  {journal} {\bibinfo  {journal} {Phys. Rev. A}\ }\textbf {\bibinfo
  {volume} {85}},\ \bibinfo {pages} {061605} (\bibinfo {year}
  {2012})}\BibitemShut {NoStop}%
\bibitem [{\citenamefont {Orth}\ \emph {et~al.}(2012)\citenamefont {Orth},
  \citenamefont {Cocks}, \citenamefont {Rachel}, \citenamefont {Buchhold},
  \citenamefont {LeHur},\ and\ \citenamefont {Hofstetter}}]{orth13}%
  \BibitemOpen
  \bibfield  {author} {\bibinfo {author} {\bibfnamefont {P.~P.}\ \bibnamefont
  {Orth}}, \bibinfo {author} {\bibfnamefont {D.}~\bibnamefont {Cocks}},
  \bibinfo {author} {\bibfnamefont {S.}~\bibnamefont {Rachel}}, \bibinfo
  {author} {\bibfnamefont {M.}~\bibnamefont {Buchhold}}, \bibinfo {author}
  {\bibfnamefont {K.}~\bibnamefont {LeHur}}, \ and\ \bibinfo {author}
  {\bibfnamefont {W.}~\bibnamefont {Hofstetter}},\ }\href@noop {} {\bibfield
  {journal} {\bibinfo  {journal} {J. Phys. B: At. Mol. Opt. Phys.}\ }\textbf
  {\bibinfo {volume} {46}},\ \bibinfo {pages} {134004} (\bibinfo {year}
  {2012})}\BibitemShut {NoStop}%
\bibitem [{\citenamefont {Grusdt}\ \emph {et~al.}(2013)\citenamefont {Grusdt},
  \citenamefont {H\"oning},\ and\ \citenamefont {Fleischhauer}}]{grusdt13}%
  \BibitemOpen
  \bibfield  {author} {\bibinfo {author} {\bibfnamefont {F.}~\bibnamefont
  {Grusdt}}, \bibinfo {author} {\bibfnamefont {M.}~\bibnamefont {H\"oning}}, \
  and\ \bibinfo {author} {\bibfnamefont {M.}~\bibnamefont {Fleischhauer}},\
  }\href@noop {} {\bibfield  {journal} {\bibinfo  {journal} {Phys. Rev. Lett.}\
  }\textbf {\bibinfo {volume} {110}},\ \bibinfo {pages} {260405} (\bibinfo
  {year} {2013})}\BibitemShut {NoStop}%
\bibitem [{\citenamefont {Grusdt}\ \emph {et~al.}(2014)\citenamefont {Grusdt},
  \citenamefont {Letscher}, \citenamefont {Hafezi},\ and\ \citenamefont
  {Fleischhauer}}]{grusdt14a}%
  \BibitemOpen
  \bibfield  {author} {\bibinfo {author} {\bibfnamefont {F.}~\bibnamefont
  {Grusdt}}, \bibinfo {author} {\bibfnamefont {F.}~\bibnamefont {Letscher}},
  \bibinfo {author} {\bibfnamefont {M.}~\bibnamefont {Hafezi}}, \ and\ \bibinfo
  {author} {\bibfnamefont {M.}~\bibnamefont {Fleischhauer}},\ }\href {\doibase
  10.1103/PhysRevLett.113.155301} {\bibfield  {journal} {\bibinfo  {journal}
  {Phys. Rev. Lett.}\ }\textbf {\bibinfo {volume} {113}},\ \bibinfo {pages}
  {155301} (\bibinfo {year} {2014})}\BibitemShut {NoStop}%
\bibitem [{\citenamefont {Grusdt}\ and\ \citenamefont
  {H\"oning}(2014)}]{grusdt14b}%
  \BibitemOpen
  \bibfield  {author} {\bibinfo {author} {\bibfnamefont {F.}~\bibnamefont
  {Grusdt}}\ and\ \bibinfo {author} {\bibfnamefont {M.}~\bibnamefont
  {H\"oning}},\ }\href {\doibase 10.1103/PhysRevA.90.053623} {\bibfield
  {journal} {\bibinfo  {journal} {Phys. Rev. A}\ }\textbf {\bibinfo {volume}
  {90}},\ \bibinfo {pages} {053623} (\bibinfo {year} {2014})}\BibitemShut
  {NoStop}%
\bibitem [{\citenamefont {Giamarchi}(2004)}]{giamarchi}%
  \BibitemOpen
  \bibfield  {author} {\bibinfo {author} {\bibfnamefont {T.}~\bibnamefont
  {Giamarchi}},\ }\href@noop {} {\emph {\bibinfo {title} {Quantum {P}hysics in
  {O}ne {D}imension}}}\ (\bibinfo  {publisher} {Clarendon Press},\ \bibinfo
  {address} {Oxford},\ \bibinfo {year} {2004})\BibitemShut {NoStop}%
\bibitem [{\citenamefont {White}(1992)}]{white92}%
  \BibitemOpen
  \bibfield  {author} {\bibinfo {author} {\bibfnamefont {S.~R.}\ \bibnamefont
  {White}},\ }\href@noop {} {\bibfield  {journal} {\bibinfo  {journal} {Phys.
  Rev. Lett.}\ }\textbf {\bibinfo {volume} {69}},\ \bibinfo {pages} {2863}
  (\bibinfo {year} {1992})}\BibitemShut {NoStop}%
\bibitem [{\citenamefont {Schollw\"ock}(2005)}]{schollwoeck05}%
  \BibitemOpen
  \bibfield  {author} {\bibinfo {author} {\bibfnamefont {U.}~\bibnamefont
  {Schollw\"ock}},\ }\href@noop {} {\bibfield  {journal} {\bibinfo  {journal}
  {Rev. Mod. Phys.}\ }\textbf {\bibinfo {volume} {77}},\ \bibinfo {pages} {259}
  (\bibinfo {year} {2005})}\BibitemShut {NoStop}%
\bibitem [{\citenamefont {Schollw\"ock}(2011)}]{schollwoeck11}%
  \BibitemOpen
  \bibfield  {author} {\bibinfo {author} {\bibfnamefont {U.}~\bibnamefont
  {Schollw\"ock}},\ }\href@noop {} {\bibfield  {journal} {\bibinfo  {journal}
  {Ann. Phys. (NY)}\ }\textbf {\bibinfo {volume} {326}},\ \bibinfo {pages} {96}
  (\bibinfo {year} {2011})}\BibitemShut {NoStop}%
\bibitem [{\citenamefont {Atala}\ \emph {et~al.}(2014)\citenamefont {Atala},
  \citenamefont {Aidelsburger}, \citenamefont {Lohse}, \citenamefont
  {Barreiro}, \citenamefont {Paredes},\ and\ \citenamefont {Bloch}}]{atala14}%
  \BibitemOpen
  \bibfield  {author} {\bibinfo {author} {\bibfnamefont {M.}~\bibnamefont
  {Atala}}, \bibinfo {author} {\bibfnamefont {M.}~\bibnamefont {Aidelsburger}},
  \bibinfo {author} {\bibfnamefont {M.}~\bibnamefont {Lohse}}, \bibinfo
  {author} {\bibfnamefont {J.~T.}\ \bibnamefont {Barreiro}}, \bibinfo {author}
  {\bibfnamefont {B.}~\bibnamefont {Paredes}}, \ and\ \bibinfo {author}
  {\bibfnamefont {I.}~\bibnamefont {Bloch}},\ }\href@noop {} {\bibfield
  {journal} {\bibinfo  {journal} {Nature Phys.}\ }\textbf {\bibinfo {volume}
  {10}},\ \bibinfo {pages} {588} (\bibinfo {year} {2014})}\BibitemShut
  {NoStop}%
\bibitem [{\citenamefont {Vekua}\ \emph {et~al.}(2003)\citenamefont {Vekua},
  \citenamefont {Japaridze},\ and\ \citenamefont {Mikeska}}]{vekua03}%
  \BibitemOpen
  \bibfield  {author} {\bibinfo {author} {\bibfnamefont {T.}~\bibnamefont
  {Vekua}}, \bibinfo {author} {\bibfnamefont {G.}~\bibnamefont {Japaridze}}, \
  and\ \bibinfo {author} {\bibfnamefont {H.-J.}\ \bibnamefont {Mikeska}},\
  }\href@noop {} {\bibfield  {journal} {\bibinfo  {journal} {Phys. Rev. B}\
  }\textbf {\bibinfo {volume} {67}},\ \bibinfo {pages} {064419} (\bibinfo
  {year} {2003})}\BibitemShut {NoStop}%
\bibitem [{\citenamefont {Cr\'epin}\ \emph {et~al.}(2011)\citenamefont
  {Cr\'epin}, \citenamefont {Laflorencie}, \citenamefont {Roux},\ and\
  \citenamefont {Simon}}]{crepin11}%
  \BibitemOpen
  \bibfield  {author} {\bibinfo {author} {\bibfnamefont {F.}~\bibnamefont
  {Cr\'epin}}, \bibinfo {author} {\bibfnamefont {N.}~\bibnamefont
  {Laflorencie}}, \bibinfo {author} {\bibfnamefont {G.}~\bibnamefont {Roux}}, \
  and\ \bibinfo {author} {\bibfnamefont {P.}~\bibnamefont {Simon}},\ }\href
  {\doibase 10.1103/PhysRevB.84.054517} {\bibfield  {journal} {\bibinfo
  {journal} {Phys. Rev. B}\ }\textbf {\bibinfo {volume} {84}},\ \bibinfo
  {pages} {054517} (\bibinfo {year} {2011})}\BibitemShut {NoStop}%
\bibitem [{\citenamefont {Orignac}\ and\ \citenamefont
  {Giamarchi}(2001)}]{orignac01}%
  \BibitemOpen
  \bibfield  {author} {\bibinfo {author} {\bibfnamefont {E.}~\bibnamefont
  {Orignac}}\ and\ \bibinfo {author} {\bibfnamefont {T.}~\bibnamefont
  {Giamarchi}},\ }\href@noop {} {\bibfield  {journal} {\bibinfo  {journal}
  {Phys. Rev. B}\ }\textbf {\bibinfo {volume} {64}},\ \bibinfo {pages} {144515}
  (\bibinfo {year} {2001})}\BibitemShut {NoStop}%
\bibitem [{\citenamefont {Cha}\ and\ \citenamefont {Shin}(2011)}]{cha11}%
  \BibitemOpen
  \bibfield  {author} {\bibinfo {author} {\bibfnamefont {M.-C.}\ \bibnamefont
  {Cha}}\ and\ \bibinfo {author} {\bibfnamefont {J.-G.}\ \bibnamefont {Shin}},\
  }\href@noop {} {\bibfield  {journal} {\bibinfo  {journal} {Phys. Rev. A}\
  }\textbf {\bibinfo {volume} {83}},\ \bibinfo {pages} {055602} (\bibinfo
  {year} {2011})}\BibitemShut {NoStop}%
\bibitem [{\citenamefont {Dhar}\ \emph {et~al.}(2012)\citenamefont {Dhar},
  \citenamefont {Maji}, \citenamefont {Mishra}, \citenamefont {Pai},
  \citenamefont {Mukerjee},\ and\ \citenamefont {Paramekanti}}]{dhar12}%
  \BibitemOpen
  \bibfield  {author} {\bibinfo {author} {\bibfnamefont {A.}~\bibnamefont
  {Dhar}}, \bibinfo {author} {\bibfnamefont {M.}~\bibnamefont {Maji}}, \bibinfo
  {author} {\bibfnamefont {T.}~\bibnamefont {Mishra}}, \bibinfo {author}
  {\bibfnamefont {R.~V.}\ \bibnamefont {Pai}}, \bibinfo {author} {\bibfnamefont
  {S.}~\bibnamefont {Mukerjee}}, \ and\ \bibinfo {author} {\bibfnamefont
  {A.}~\bibnamefont {Paramekanti}},\ }\href@noop {} {\bibfield  {journal}
  {\bibinfo  {journal} {Phys. Rev. A}\ }\textbf {\bibinfo {volume} {85}},\
  \bibinfo {pages} {041602} (\bibinfo {year} {2012})}\BibitemShut {NoStop}%
\bibitem [{\citenamefont {Dhar}\ \emph {et~al.}(2013)\citenamefont {Dhar},
  \citenamefont {Mishra}, \citenamefont {Maji}, \citenamefont {Pai},
  \citenamefont {Mukerjee},\ and\ \citenamefont {Paramekanti}}]{dhar13}%
  \BibitemOpen
  \bibfield  {author} {\bibinfo {author} {\bibfnamefont {A.}~\bibnamefont
  {Dhar}}, \bibinfo {author} {\bibfnamefont {T.}~\bibnamefont {Mishra}},
  \bibinfo {author} {\bibfnamefont {M.}~\bibnamefont {Maji}}, \bibinfo {author}
  {\bibfnamefont {R.~V.}\ \bibnamefont {Pai}}, \bibinfo {author} {\bibfnamefont
  {S.}~\bibnamefont {Mukerjee}}, \ and\ \bibinfo {author} {\bibfnamefont
  {A.}~\bibnamefont {Paramekanti}},\ }\href@noop {} {\bibfield  {journal}
  {\bibinfo  {journal} {Phys. Rev. B}\ }\textbf {\bibinfo {volume} {87}},\
  \bibinfo {pages} {174501} (\bibinfo {year} {2013})}\BibitemShut {NoStop}%
\bibitem [{\citenamefont {Petrescu}\ and\ \citenamefont
  {Le~Hur}(2013)}]{petrescu13}%
  \BibitemOpen
  \bibfield  {author} {\bibinfo {author} {\bibfnamefont {A.}~\bibnamefont
  {Petrescu}}\ and\ \bibinfo {author} {\bibfnamefont {K.}~\bibnamefont
  {Le~Hur}},\ }\href@noop {} {\bibfield  {journal} {\bibinfo  {journal} {Phys.
  Rev. Lett.}\ }\textbf {\bibinfo {volume} {111}},\ \bibinfo {pages} {150601}
  (\bibinfo {year} {2013})}\BibitemShut {NoStop}%
\bibitem [{\citenamefont {H\"ugel}\ and\ \citenamefont
  {Paredes}(2014)}]{huegel14}%
  \BibitemOpen
  \bibfield  {author} {\bibinfo {author} {\bibfnamefont {D.}~\bibnamefont
  {H\"ugel}}\ and\ \bibinfo {author} {\bibfnamefont {B.}~\bibnamefont
  {Paredes}},\ }\href@noop {} {\bibfield  {journal} {\bibinfo  {journal} {Phys.
  Rev. A}\ }\textbf {\bibinfo {volume} {89}},\ \bibinfo {pages} {023619}
  (\bibinfo {year} {2014})}\BibitemShut {NoStop}%
\bibitem [{\citenamefont {Wei}\ and\ \citenamefont {Mueller}(2014)}]{wei14}%
  \BibitemOpen
  \bibfield  {author} {\bibinfo {author} {\bibfnamefont {R.}~\bibnamefont
  {Wei}}\ and\ \bibinfo {author} {\bibfnamefont {E.~J.}\ \bibnamefont
  {Mueller}},\ }\href@noop {} {\bibfield  {journal} {\bibinfo  {journal} {Phys.
  Rev. A}\ }\textbf {\bibinfo {volume} {89}},\ \bibinfo {pages} {063617}
  (\bibinfo {year} {2014})}\BibitemShut {NoStop}%
\bibitem [{\citenamefont {Tokuno}\ and\ \citenamefont
  {Georges}(2014)}]{tokuno14}%
  \BibitemOpen
  \bibfield  {author} {\bibinfo {author} {\bibfnamefont {A.}~\bibnamefont
  {Tokuno}}\ and\ \bibinfo {author} {\bibfnamefont {A.}~\bibnamefont
  {Georges}},\ }\href@noop {} {\bibfield  {journal} {\bibinfo  {journal} {New
  J. Phys.}\ }\textbf {\bibinfo {volume} {16}},\ \bibinfo {pages} {073005}
  (\bibinfo {year} {2014})}\BibitemShut {NoStop}%
\bibitem [{\citenamefont {Lim}\ \emph {et~al.}(2008)\citenamefont {Lim},
  \citenamefont {Smith},\ and\ \citenamefont {Hemmerich}}]{lim08}%
  \BibitemOpen
  \bibfield  {author} {\bibinfo {author} {\bibfnamefont {L.-K.}\ \bibnamefont
  {Lim}}, \bibinfo {author} {\bibfnamefont {C.}~\bibnamefont {Smith}}, \ and\
  \bibinfo {author} {\bibfnamefont {A.}~\bibnamefont {Hemmerich}},\ }\href
  {\doibase 10.1103/PhysRevLett.100.130402} {\bibfield  {journal} {\bibinfo
  {journal} {Phys. Rev. Lett.}\ }\textbf {\bibinfo {volume} {100}},\ \bibinfo
  {pages} {130402} (\bibinfo {year} {2008})}\BibitemShut {NoStop}%
\bibitem [{\citenamefont {M\"oller}\ and\ \citenamefont
  {Cooper}(2010)}]{moeller10}%
  \BibitemOpen
  \bibfield  {author} {\bibinfo {author} {\bibfnamefont {G.}~\bibnamefont
  {M\"oller}}\ and\ \bibinfo {author} {\bibfnamefont {N.}~\bibnamefont
  {Cooper}},\ }\href {\doibase 10.1103/PhysRevA.82.063625} {\bibfield
  {journal} {\bibinfo  {journal} {Phys. Rev. A}\ }\textbf {\bibinfo {volume}
  {82}},\ \bibinfo {pages} {063625} (\bibinfo {year} {2010})}\BibitemShut
  {NoStop}%
\bibitem [{sup()}]{suppmat}%
  \BibitemOpen
  \href@noop {} {}\bibinfo {note} {See Supplemental Material}\BibitemShut
  {NoStop}%
\bibitem [{\citenamefont {Arlego}\ \emph {et~al.}(2011)\citenamefont {Arlego},
  \citenamefont {Heidrich-Meisner}, \citenamefont {Honecker}, \citenamefont
  {Rossini},\ and\ \citenamefont {Vekua}}]{arlego12}%
  \BibitemOpen
  \bibfield  {author} {\bibinfo {author} {\bibfnamefont {M.}~\bibnamefont
  {Arlego}}, \bibinfo {author} {\bibfnamefont {F.}~\bibnamefont
  {Heidrich-Meisner}}, \bibinfo {author} {\bibfnamefont {A.}~\bibnamefont
  {Honecker}}, \bibinfo {author} {\bibfnamefont {G.}~\bibnamefont {Rossini}}, \
  and\ \bibinfo {author} {\bibfnamefont {T.}~\bibnamefont {Vekua}},\
  }\href@noop {} {\bibfield  {journal} {\bibinfo  {journal} {Phys. Rev. B}\
  }\textbf {\bibinfo {volume} {84}},\ \bibinfo {pages} {224409} (\bibinfo
  {year} {2011})}\BibitemShut {NoStop}%
\bibitem [{\citenamefont {Kolezhuk}\ \emph {et~al.}(2012)\citenamefont
  {Kolezhuk}, \citenamefont {Heidrich-Meisner}, \citenamefont {Greschner},\
  and\ \citenamefont {Vekua}}]{kolezhuk12}%
  \BibitemOpen
  \bibfield  {author} {\bibinfo {author} {\bibfnamefont {A.~K.}\ \bibnamefont
  {Kolezhuk}}, \bibinfo {author} {\bibfnamefont {F.}~\bibnamefont
  {Heidrich-Meisner}}, \bibinfo {author} {\bibfnamefont {S.}~\bibnamefont
  {Greschner}}, \ and\ \bibinfo {author} {\bibfnamefont {T.}~\bibnamefont
  {Vekua}},\ }\href@noop {} {\bibfield  {journal} {\bibinfo  {journal} {Phys.
  Rev. B}\ }\textbf {\bibinfo {volume} {85}},\ \bibinfo {pages} {064420}
  (\bibinfo {year} {2012})}\BibitemShut {NoStop}%
\bibitem [{\citenamefont {Shyiko}\ \emph {et~al.}(2013)\citenamefont {Shyiko},
  \citenamefont {McCulloch}, \citenamefont {Gumenjuk-Sichevska},\ and\
  \citenamefont {Kolezhuk}}]{shyiko13}%
  \BibitemOpen
  \bibfield  {author} {\bibinfo {author} {\bibfnamefont {I.~T.}\ \bibnamefont
  {Shyiko}}, \bibinfo {author} {\bibfnamefont {I.~P.}\ \bibnamefont
  {McCulloch}}, \bibinfo {author} {\bibfnamefont {J.~V.}\ \bibnamefont
  {Gumenjuk-Sichevska}}, \ and\ \bibinfo {author} {\bibfnamefont {A.~K.}\
  \bibnamefont {Kolezhuk}},\ }\href {\doibase 10.1103/PhysRevB.88.014403}
  {\bibfield  {journal} {\bibinfo  {journal} {Phys. Rev. B}\ }\textbf {\bibinfo
  {volume} {88}},\ \bibinfo {pages} {014403} (\bibinfo {year}
  {2013})}\BibitemShut {NoStop}%
\bibitem [{\citenamefont {Bloch}\ \emph {et~al.}(2008)\citenamefont {Bloch},
  \citenamefont {Dalibard},\ and\ \citenamefont {Zwerger}}]{bloch08}%
  \BibitemOpen
  \bibfield  {author} {\bibinfo {author} {\bibfnamefont {I.}~\bibnamefont
  {Bloch}}, \bibinfo {author} {\bibfnamefont {J.}~\bibnamefont {Dalibard}}, \
  and\ \bibinfo {author} {\bibfnamefont {W.}~\bibnamefont {Zwerger}},\
  }\href@noop {} {\bibfield  {journal} {\bibinfo  {journal} {Rev. Mod. Phys.}\
  }\textbf {\bibinfo {volume} {80}},\ \bibinfo {pages} {885} (\bibinfo {year}
  {2008})}\BibitemShut {NoStop}%
\bibitem [{\citenamefont {Mancini}\ \emph {et~al.}(shed)\citenamefont
  {Mancini}, \citenamefont {Pagano}, \citenamefont {Cappellini}, \citenamefont
  {Livi}, \citenamefont {Rider}, \citenamefont {Catani}, \citenamefont {Sias},
  \citenamefont {Zoller}, \citenamefont {Inguscio}, \citenamefont {Dalmonte},\
  and\ \citenamefont {Fallani}}]{mancini15}%
  \BibitemOpen
  \bibfield  {author} {\bibinfo {author} {\bibfnamefont {M.}~\bibnamefont
  {Mancini}}, \bibinfo {author} {\bibfnamefont {G.}~\bibnamefont {Pagano}},
  \bibinfo {author} {\bibfnamefont {G.}~\bibnamefont {Cappellini}}, \bibinfo
  {author} {\bibfnamefont {L.}~\bibnamefont {Livi}}, \bibinfo {author}
  {\bibfnamefont {M.}~\bibnamefont {Rider}}, \bibinfo {author} {\bibfnamefont
  {J.}~\bibnamefont {Catani}}, \bibinfo {author} {\bibfnamefont
  {C.}~\bibnamefont {Sias}}, \bibinfo {author} {\bibfnamefont {P.}~\bibnamefont
  {Zoller}}, \bibinfo {author} {\bibfnamefont {M.}~\bibnamefont {Inguscio}},
  \bibinfo {author} {\bibfnamefont {M.}~\bibnamefont {Dalmonte}}, \ and\
  \bibinfo {author} {\bibfnamefont {L.}~\bibnamefont {Fallani}},\ }\href@noop
  {} {\ ,\ \bibinfo {pages} {arXiv:1502.02495} (\bibinfo {year}
  {unpublished})}\BibitemShut {NoStop}%
\bibitem [{\citenamefont {Stuhl}\ \emph {et~al.}(shed)\citenamefont {Stuhl},
  \citenamefont {Lu}, \citenamefont {Aycock}, \citenamefont {Genkina},\ and\
  \citenamefont {Spielman}}]{leblanc15}%
  \BibitemOpen
  \bibfield  {author} {\bibinfo {author} {\bibfnamefont {B.~K.}\ \bibnamefont
  {Stuhl}}, \bibinfo {author} {\bibfnamefont {H.-I.}\ \bibnamefont {Lu}},
  \bibinfo {author} {\bibfnamefont {L.~M.}\ \bibnamefont {Aycock}}, \bibinfo
  {author} {\bibfnamefont {D.}~\bibnamefont {Genkina}}, \ and\ \bibinfo
  {author} {\bibfnamefont {I.~B.}\ \bibnamefont {Spielman}},\ }\href@noop {} {\
  ,\ \bibinfo {pages} {arXiv:1502.02496} (\bibinfo {year}
  {unpublished})}\BibitemShut {NoStop}%
\bibitem [{\citenamefont {Vekua}\ \emph {et~al.}(2004)\citenamefont {Vekua},
  \citenamefont {Japaridze},\ and\ \citenamefont {Mikeska}}]{vekua04}%
  \BibitemOpen
  \bibfield  {author} {\bibinfo {author} {\bibfnamefont {T.}~\bibnamefont
  {Vekua}}, \bibinfo {author} {\bibfnamefont {G.~I.}\ \bibnamefont
  {Japaridze}}, \ and\ \bibinfo {author} {\bibfnamefont {H.-J.}\ \bibnamefont
  {Mikeska}},\ }\href@noop {} {\bibfield  {journal} {\bibinfo  {journal} {Phys.
  Rev. B}\ }\textbf {\bibinfo {volume} {70}},\ \bibinfo {pages} {014425}
  (\bibinfo {year} {2004})}\BibitemShut {NoStop}%
\bibitem [{\citenamefont {Fendley}\ \emph {et~al.}(1993)\citenamefont
  {Fendley}, \citenamefont {Saleur},\ and\ \citenamefont
  {Zamolodchikov}}]{Zamolodchikov93}%
  \BibitemOpen
  \bibfield  {author} {\bibinfo {author} {\bibfnamefont {P.}~\bibnamefont
  {Fendley}}, \bibinfo {author} {\bibfnamefont {H.}~\bibnamefont {Saleur}}, \
  and\ \bibinfo {author} {\bibfnamefont {A.~B.}\ \bibnamefont
  {Zamolodchikov}},\ }\href@noop {} {\bibfield  {journal} {\bibinfo  {journal}
  {Int. J. Mod. Phys. A}\ }\textbf {\bibinfo {volume} {8}},\ \bibinfo {pages}
  {5751} (\bibinfo {year} {1993})}\BibitemShut {NoStop}%
\bibitem [{\citenamefont {Zamolodchikov}(1995)}]{Zamolodchikov95}%
  \BibitemOpen
  \bibfield  {author} {\bibinfo {author} {\bibfnamefont {A.~B.}\ \bibnamefont
  {Zamolodchikov}},\ }\href@noop {} {\bibfield  {journal} {\bibinfo  {journal}
  {Int. J. Mod. Phys. A}\ }\textbf {\bibinfo {volume} {10}},\ \bibinfo {pages}
  {1125} (\bibinfo {year} {1995})}\BibitemShut {NoStop}%
\bibitem [{\citenamefont {Okunishi}\ and\ \citenamefont
  {Tonegawa}(2003)}]{okunishi03a}%
  \BibitemOpen
  \bibfield  {author} {\bibinfo {author} {\bibfnamefont {K.}~\bibnamefont
  {Okunishi}}\ and\ \bibinfo {author} {\bibfnamefont {T.}~\bibnamefont
  {Tonegawa}},\ }\href@noop {} {\bibfield  {journal} {\bibinfo  {journal}
  {Phys. Rev. B}\ }\textbf {\bibinfo {volume} {68}},\ \bibinfo {pages} {224422}
  (\bibinfo {year} {2003})}\BibitemShut {NoStop}%
\bibitem [{\citenamefont {Vidal}\ \emph {et~al.}(2003)\citenamefont {Vidal},
  \citenamefont {Latorre}, \citenamefont {Rico},\ and\ \citenamefont
  {Kitaev}}]{Vidal03}%
  \BibitemOpen
  \bibfield  {author} {\bibinfo {author} {\bibfnamefont {G.}~\bibnamefont
  {Vidal}}, \bibinfo {author} {\bibfnamefont {J.~I.}\ \bibnamefont {Latorre}},
  \bibinfo {author} {\bibfnamefont {E.}~\bibnamefont {Rico}}, \ and\ \bibinfo
  {author} {\bibfnamefont {A.}~\bibnamefont {Kitaev}},\ }\href@noop {}
  {\bibfield  {journal} {\bibinfo  {journal} {Phys. Rev. Lett.}\ }\textbf
  {\bibinfo {volume} {90}},\ \bibinfo {pages} {227902} (\bibinfo {year}
  {2003})}\BibitemShut {NoStop}%
\bibitem [{\citenamefont {Calabrese}\ and\ \citenamefont
  {J.~Cardy}(2004)}]{Calabrese04}%
  \BibitemOpen
  \bibfield  {author} {\bibinfo {author} {\bibfnamefont {P.}~\bibnamefont
  {Calabrese}}\ and\ \bibinfo {author} {\bibfnamefont {J.}~\bibnamefont
  {J.~Cardy}},\ }\href@noop {} {\bibfield  {journal} {\bibinfo  {journal} {J.
  Stat. Mech.: Theory Exp.}\ ,\ \bibinfo {pages} {P06002}} (\bibinfo {year}
  {2004})}\BibitemShut {NoStop}%
\end{thebibliography}
\end{document}